\documentclass{aa}  

\usepackage{graphicx}
\usepackage{txfonts}
\usepackage[usenames]{color} 
\usepackage{tabularx} 
\usepackage{footnote}
\makesavenoteenv{table*}
\usepackage{scalerel}
\usepackage{tikz}
\usetikzlibrary{svg.path}
\definecolor{orcidlogocol}{HTML}{A6CE39}
\tikzset{
  orcidlogo/.pic={
    \fill[orcidlogocol] svg{M256,128c0,70.7-57.3,128-128,128C57.3,256,0,198.7,0,128C0,57.3,57.3,0,128,0C198.7,0,256,57.3,256,128z};
    \fill[white] svg{M86.3,186.2H70.9V79.1h15.4v48.4V186.2z}
                 svg{M108.9,79.1h41.6c39.6,0,57,28.3,57,53.6c0,27.5-21.5,53.6-56.8,53.6h-41.8V79.1z M124.3,172.4h24.5c34.9,0,42.9-26.5,42.9-39.7c0-21.5-13.7-39.7-43.7-39.7h-23.7V172.4z}
                 svg{M88.7,56.8c0,5.5-4.5,10.1-10.1,10.1c-5.6,0-10.1-4.6-10.1-10.1c0-5.6,4.5-10.1,10.1-10.1C84.2,46.7,88.7,51.3,88.7,56.8z};
  }
}

\newcommand\orcid[1]{\href{https://orcid.org/#1}{\mbox{\scalerel*{
\begin{tikzpicture}[yscale=-1,transform shape]
\pic{orcidlogo};
\end{tikzpicture}
}{|}}} \href{#1}{#1}}
\usepackage[breaklinks=true,colorlinks, linkcolor=blue,urlcolor=blue,citecolor=blue]{hyperref}

\usepackage{ulem}
 
\begin{document}

   \title{Evidence of a complex structure within the 2013 August 19 coronal mass ejection}

   \subtitle{Radial and longitudinal evolution in the inner heliosphere}

   \author{L. Rodríguez-García
          \inst{1}\and T. Nieves-Chinchilla\inst{2} \and R. Gómez-Herrero\inst{1}
          \and I. Zouganelis\inst{3}
          \and A. Vourlidas\inst{4} \and L. A. Balmaceda\inst{2,5}  
           \and \newline  M. Dumbovi\'c\inst{6} \and  L. K. Jian\inst{2}  \and L. Mays\inst{2}  \and  F. Carcaboso\inst{1,2,7} \and L. F. G. dos Santos\inst{8} \and
           J. Rodríguez-Pacheco\inst{1}
          }
   \institute{Universidad de Alcalá, Space Research Group, Alcalá de Henares, Madrid, Spain \\
              \email{l.rodriguezgarcia@edu.uah.es}
             \and
            Heliophysics Science Division, NASA Goddard Space Flight Center, Greenbelt, MD, USA
            \and
             European Space Agency (ESA), European Space Astronomy Centre (ESAC), Camino Bajo del Castillo s/n, 28692 Villanueva de la Cañada, Madrid, Spain
             \and
              Johns Hopkins University Applied Physics Laboratory, Laurel, MD, USA
            \and 
             George Mason University, Fairfax, VA, USA
            \and
             Hvar Observatory, Faculty of Geodesy, University of Zagreb, Croatia
             \and
            The Catholic University of America, Washington, DC, USA
            \and
            Cooperative Institute for Research in Environmental Sciences, University of Colorado Boulder, Boulder, CO, USA
            }

 
  \abstract
  {Late on 2013 August 19, a coronal mass ejection (CME) erupted from an active region located near the far-side central meridian from Earth's perspective. The event and its accompanying shock were remotely observed by the STEREO-A, STEREO-B and SOHO spacecraft. The interplanetary counterpart (ICME) was intercepted by MESSENGER near 0.3 au, and by both STEREO-A and STEREO-B, near 1 au, which were separated by 78$^{\circ}$ in heliolongitude.}
   {The main objective of this study is to follow the radial and longitudinal evolution of the ICME throughout the inner heliosphere, and to examine  possible scenarios for the different magnetic flux-rope configuration observed on the solar disk, and measured in situ at the locations of MESSENGER and STEREO-A, separated by 15$^{\circ}$ in heliolongitude, and at STEREO-B, which detected the ICME flank.} 
   {Solar disk observations are used to estimate the `magnetic flux-rope type', namely, the magnetic helicity, axis orientation and axial magnetic field direction of the flux rope. The graduated cylindrical shell model is used to reconstruct the CME in the corona. The analysis of in-situ data, specifically, plasma and magnetic field, is used to estimate the global interplanetary shock geometry and to derive the magnetic flux-rope type at different in-situ locations, which is compared to the type estimated from solar disk observations. The elliptical cylindrical analytical model is used for the in-situ magnetic flux-rope reconstruction.}
   {Based on the CME geometry and on the spacecraft configuration, we find that the magnetic flux-rope structure detected at STEREO-B belongs to the same ICME detected at MESSENGER and STEREO-A. The opposite helicity deduced at STEREO-B, might be due to the spacecraft intercepting one of the legs of the structure far from the flux-rope axis, while STEREO-A and MESSENGER are crossing through the core of the magnetic flux rope. The different flux-rope orientations measured at MESSENGER and STEREO-A arise probably because the two spacecraft measure a curved, highly distorted and rather complex magnetic flux-rope topology. The ICME may have suffered additional distortion in its evolution in the inner heliosphere, such as the west flank is propagating faster than the east flank when arriving near 1 au. }
   {This work illustrates how a wide, curved, highly distorted and rather complex CME was showing different  orientations as observed on the solar disk and measured in situ at 0.3 au and near 1 au. It shows how the ambient conditions can significantly affect the expansion and propagation of the CME/ICME, introducing additional irregularities to the already asymmetric eruption. The study also manifests how these complex structures cannot be directly reconstructed with the current models available, and that multi-point analysis is of the utmost importance in such complex events. }
    \maketitle
%
\section{Introduction}
\label{sec:Introduc}
Coronal mass ejections (CMEs) are large eruptions of magnetized plasma that are expelled from the Sun into the heliosphere as a result of the release of the huge energy stored in the solar magnetic field structures. Remote-sensing observations of CMEs close to the Sun provide evidence for the existence of magnetic flux-rope (MFR) structures within CMEs \citep{vourlidas_flux_2014}. They consist of confined plasma within a helically organized magnetic structure, and its general pattern can be figured out in reference to the `MFR type' \citep[e.g., South-East-North,  West-North-East, etc., as in][]{BothmerSchwenn1998,MulliganRussellLuhmann1998}. To estimate the MFR type, the orientation of the MFR axis with respect to the solar equator (tilt angle), the direction of its axial magnetic field, and its magnetic helicity sign or chirality can be inferred from remote-sensing solar disk observations \citep{Palmerio2017}. The 3D MFR geometry, in terms of CME propagation direction, orientation, width and speed, can be derived from the imaging observations from multiple vantage points, to minimize the projection effects, using a forward modeling, such as the graduated cylindrical shell model \citep[GCS,][]{Thernisien2006GCS,Thernisien2011}. 

In the interplanetary (IP) space, the evidence of MFR are found in structures known as magnetic clouds (MCs) \citep{Burlaga1981}. In the best examples, the in-situ MFR signatures are consistent and simultaneous with the plasma parameters behaviour in the defined MC structure: a monotonic rotation of the magnetic field direction through a large angle is seen along with low plasma temperature and low plasma $\beta$.
There are several models built for reconstructing MCs, such as the concept of a flux rope in a force-free configuration \citep{Burlaga1988,Lepping1990}, or those relaxing the force-free conditions \citep[e.g.,][]{Owens2006}. \cite{2018aNievesChinchilla} developed the elliptical cylindrical analytical MFR model for MCs (hereafter the EC model) as an approach to consider the distorted cross-section of the magnetic field topology, as a possible effect for the MFR interaction with the solar wind. However, all the models describe a limited subset of the properties of a MC, as they are based on one-dimensional measurements along a line cutting through the structure. 
  
Generally, the observed magnetic field signatures are more complex than a simple MFR magnetic topology \citep{Burlaga2002,RichardsonCane2004a}. To explain these partial MFR signatures, \cite{Jian2006} extended the MC definition by adopting the concept of the magnetic obstacle (MO), to which \cite{Nieves-Chinchilla2018DiP} later added the criterion of low plasma $\beta$. In addition, MCs are not always detected within the CME IP counterparts, hereafter IP CMEs (ICMEs) \citep[][]{ZurbuchenRichardson2006,RichardsonCane2010}. This might result from spacecraft crossing far from the ICME core \citep[e.g.,][]{CaneRichardsonWibberenz1997,Jian2006,Kilpua2011}, or from the erosion of the magnetic structure close to the Sun or in the IP medium \citep[e.g.,][]{Dasso2007,2016Winslow,2021Dumbovic}. 

Observations also suggest that the CME, and presumably the MFR, undergo significant evolutionary changes, such as rotation, deformation, or distortion during their heliospheric propagation, among others, \citep{Burlaga2002,Nieves-Chinchilla2012, Manchester2017, Florido2020}, with a small number of ICMEs evolving in the shape of outward-concave arcs \citep{Kahler2007}. To better understand the effects of the evolutionary processes in the CME, the in-situ measurements are compared with the remote-sensing observations. Effects, such as rotation or expansion, can be captured by the forward modeling techniques (e.g., GCS model). The MFR type can be classified based on the different magnetic configurations of MCs and their magnetic helicity, left-handed (LH) or right-handed (RH), based on the magnetic flux tube concept and the field rotation that a spacecraft would observe during the cloud's passage \citep{BothmerSchwenn1998}. Then, the MFR type estimated from in-situ measurements can be compared to the MFR type derived from solar disk observations.

The radial and longitudinal evolution of the ICMEs can be investigated thanks to constellations of spacecraft distributed throughout the heliosphere, such as the \textit{MErcury Surface Space ENvironment GEochemistry and Ranging} \citep[MESSENGER,][]{Solomon2007MESSENGER}, located near 0.3 au; the \textit{SOlar and Heliographic Observatory} \citep[SOHO,][]{Domingo1995SOHO}, and the \textit{Solar TErrestrial RElations Observatory} \citep[STEREO,][]{Kaiser2008STEREO}, located near 1 au. The ever-changing configuration of the STEREO spacecraft is advantageous for the study of the longitudinal extent of ICMEs near 1 au. It is also known that at solar maximum ICMEs are wider and with higher magnetic field, so they can be easily encountered by spacecraft separated by several tens of degrees \citep{Jian2006}.

In this study, we analyzed the evolution of the ICME associated with a CME that erupted on 2013 August 19, near the far-side central meridian from Earth's perspective. This wide and fast CME was related with an unusual widespread solar energetic particle (SEP) event analysed in detail by \cite{2021Rodriguez-Garcia}, hereafter Paper I. There were discrepancies between the MFR types as observed on the solar disk and measured in situ at the locations of MESSENGER, STEREO-A, and STEREO-B, spanning a longitudinal range of 78$^{\circ}$ in the ecliptic plane. Assuming that the ICME intercepted by the three spacecraft was the same, the ICME seemed to suffer from distortion in its evolution in the heliosphere, such as the west flank was propagating faster than the east flank when arriving near 1 au. In addition, the ICME showed a complex magnetic structure measured at the locations of MESSENGER, at 0.33 au, and STEREO-A, near 1 au, separated by 15$^{\circ}$ in heliolongitude.

The goals of this study are (1) to determine if the ICME observed by the three spacecraft is the same and (2) to better understand the configuration, geometry and causes of the MFR type discrepancies among the spacecraft. To accomplish these goals, we analyzed the remote-sensing and in-situ observations presented in Sect. \ref{sec:OBSERV}. We modelled the CME using GCS analysis and derived the MFR type based on the chirality, axis orientation and axial magnetic field direction from solar disk observations in Sect. \ref{sec:Remote-sensing analysis}. We examined in-situ measurements in Sect. \ref{sec:in situ analysis} to extrapolate the global shape of the ICME, including the global IP shock surface, and the MFR types at different locations. The EC model was used to reconstruct the MCs. The reconciliation between remote-sensing and in-situ observations is presented in Sect. \ref{sec:remote in situ reconciliation}. Section \ref{sec:summary and discussion} summarizes the observations and analysis and includes the discussion. Finally, Sect. \ref{sec:conclusions} outlines the main conclusions. The instrumentation and catalogues used in this study are introduced in the following section.

\begin{figure*}
\centering
  \resizebox{\hsize}{!}{\includegraphics{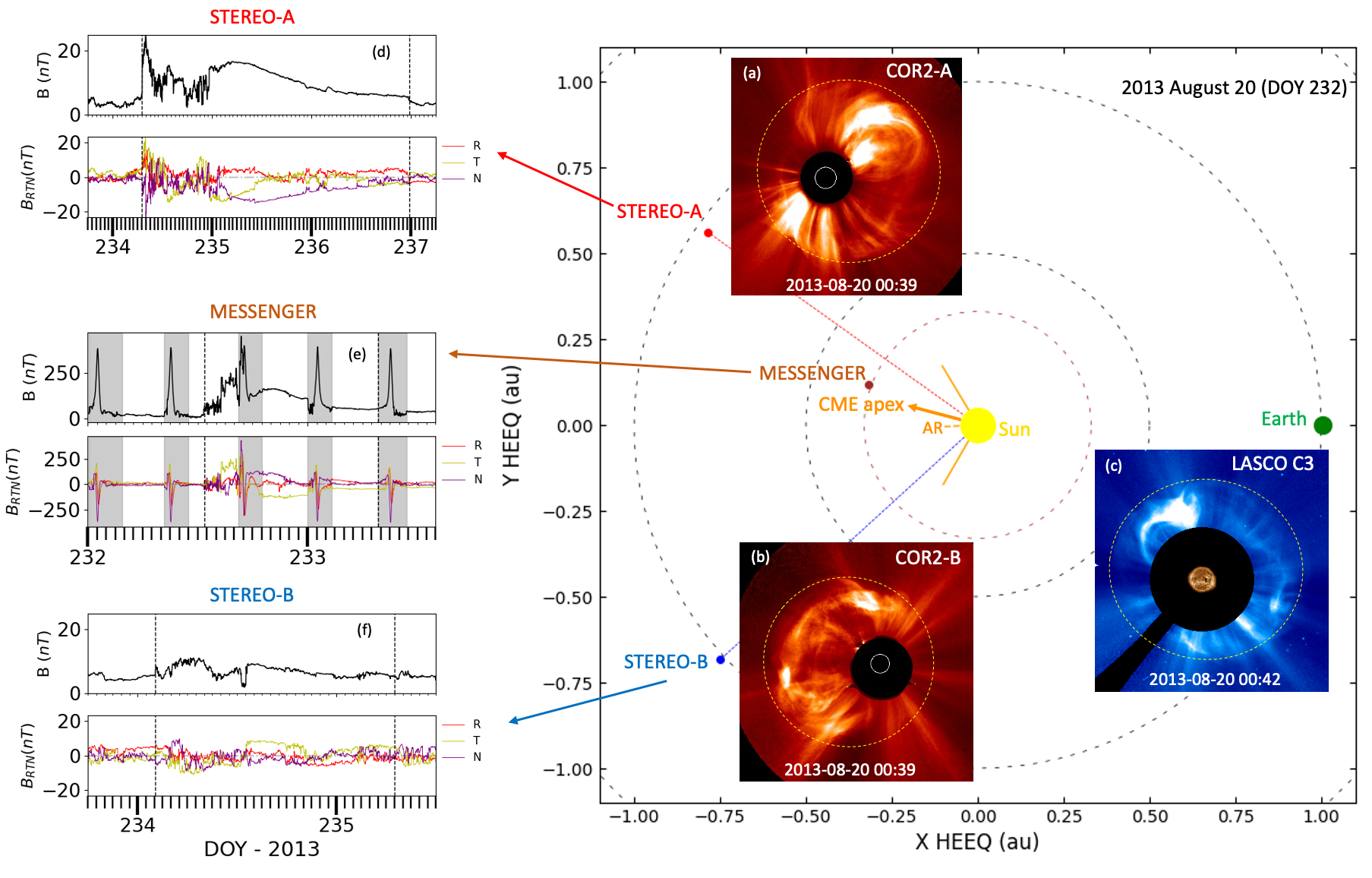}}
     \caption{ Remote-sensing observations and in-situ measurements of the 2013 August 19 CME. \textit{Right}: View from the north heliographic pole showing the positions of STEREO-A (W144), MESSENGER (W159), STEREO-B (E138) and the Earth on 2013 August 20 (DOY 232) in HEEQ coordinates. The yellow circle at the center represents the Sun (not to scale). The orange dashed line marks the active region (AR) longitude. The orange arrow indicates the longitude of the CME apex direction (W165), where the orange lines (W121, E120) outline the width of the CME, as derived from the GCS model. The dashed black circles correspond to heliocentric distances of 0.5 R\textsubscript{$\odot$} and 1 R\textsubscript{$\odot$}, respectively, while the brown circle corresponds to the heliocentric distance of MESSENGER (0.33 au). The inserted pictures show the coronagraph images as seen by COR2-A (a), COR2-B (b), and C3 (c), where the images have been magnified for a better visualization. The yellow dashed curves roughly outlined the CME-driven shock. \textit{Left}: The solar wind magnetic field measured by STEREO-A (d), MESSENGER (e) and STEREO-B (f). The \textit{top panels} show the magnetic field strength and the \textit{bottom panels} show the magnetic field B\textsubscript{RTN} components, where RTN corresponds to Radial-Tangential-Normal coordinate system \cite[e.g.,][]{Hapgood1992}. The dashed vertical lines mark the starting and ending of the ICME. }
     \label{fig:Orbit_solar_wind}
\end{figure*}

\section{Instrumentation and catalogues}
\label{sec:Instrumentation}
The study of the global geometry and configuration of the CME/ICME requires the analysis of observations from different instrumentation on board several spacecraft. We used data from STEREO-A, STEREO-B, MESSENGER, SOHO, and the \textit{Solar Dynamics Observatory} \citep[SDO,][]{Pesnell2012}. Remote-sensing observations of the CME and the activity on the solar surface were provided by the Helioseismic and Magnetic Imager \citep[HMI,][]{2012ScherrerHMI} on board SDO, the C2 and C3 coronagraphs of the Large Angle and Spectrometric COronagraph \citep[LASCO,][]{Brueckner1995} instrument on board SOHO, and the Sun Earth Connection Coronal and Heliospheric Investigation \citep[SECCHI,][]{Howard2008SECCHI} instrument suite on board STEREO. In particular, the COR1 and COR2 coronagraphs and the Extreme Ultraviolet Imager \citep[EUVI,][]{Wuelser2004}, part of SECCHI suite, were utilized. Solar wind plasma and magnetic field observations were obtained from the Plasma and Suprathermal Ion Composition \citep[PLASTIC,][]{Galvin2008} investigation and the Magnetic Field Experiment \citep{Acuna2008} on board STEREO; and the Magnetometer Instrument \citep{Anderson2007} on board MESSENGER.

Interplanetary disturbances were consulted using the STEREO level 3 event lists\footnote{\url{https://stereo-ssc.nascom.nasa.gov/data/ins_data/impact/level3/}\label{footnote stereo_level3}} \citep{Jian2018,Jian2019} maintained by L. Jian, the near-Earth ICME list provided by I. Richardson and H. Cane\footnote{\url{http://www.srl.caltech.edu/ACE/ASC/DATA/level3/icmetable2.htm}\label{footnote Richardson list}} \citep{RichardsonCane2010}, the IP shocks catalogue maintained by the University of Helsinki\footnote{\url{http://www.ipshocks.fi/}\label{footnote IP Helsinki}}, and the ICME catalogue at Mercury from the University of New Hampshire\footnote{\url{http://c-swepa.sr.unh.edu/icmecatalogatmercury.html}\label{footnote ICME Mercury list}} \citep{2015Winslow}.

\section{Observations of the 2013 August 19 CME}
\label{sec:OBSERV}
Late on 2013 August 19 (DOY 231), a CME erupting close to the far-side central meridian as seen from Earth's perspective was observed by both STEREO/SECCHI and SOHO/LASCO telescopes. The associated ICME was measured in situ by MESSENGER, STEREO-A and presumably also by STEREO-B, between 2013 DOY 232 and DOY 237.

Right panel of Fig. \ref{fig:Orbit_solar_wind} displays the map of the locations of the observatories and a summary of the remote observations of the 2013 August 19 CME. The map in Heliocentric Earth equatorial (HEEQ) coordinates shows the locations of STEREO-A, STEREO-B and MESSENGER on DOY 232, with respect to the near-Earth spacecraft, specifically SOHO, as viewed from the north heliographic pole. On the map, the coronagraph images showing the CME are inserted next to the location of the respective observing spacecraft. At the time of the eruption, STEREO-A and STEREO-B were respectively located $\sim$144$^{\circ}$ west and $\sim$138$^{\circ}$ east from Earth. The STEREO spacecraft were separated by $\sim$78$^{\circ}$ in heliolongitude, while MESSENGER and STEREO-A were separated by $\sim$15$^{\circ}$.

Figures \ref{fig:Orbit_solar_wind}d--f respectively show the magnetic field magnitude and components measured by STEREO-A, MESSENGER and STEREO-B. The ICME, starting with the arrival of the IP shock, was observed first by the Magnetometer Instrument on board MESSENGER (e), located at 0.33 au, secondly by the Magnetic Field Experiment on board STEREO-B (f), located at 1.02 au, and lastly by the Magnetic Field Experiment on board STEREO-A (d), located at 0.97 au. The magnetic field data show that both MESSENGER and STEREO-A respectively display a coherent change in the magnetic field during $\sim$12 hours and $\sim$48 hours, with an increase in the magnitude and a smooth rotation in the magnetic field components. Both spacecraft observe similar magnetic field configuration but one of the components depicts different polarity. The B\textsubscript{N} is positive in MESSENGER and negative in STEREO-A. STEREO-B also observes a rotation in the magnetic field components but only during $\sim$8 hours. Although the latitudinal distance between the STEREO spacecraft was very small ($\sim$2$^{\circ}$), given the longitudinal separation between them, it is not straightforward to evaluate if the structure observed by STEREO-B is the same to the one observed by MESSENGER and STEREO-A and it requires an analysis that 
will be performed in Sect. \ref{sec:in situ analysis}. 

\subsection{Remote-sensing observations}
\label{sec:OBSER_remote-sensing}

\begin{figure*}[htbp]
 \includegraphics[width=18cm]{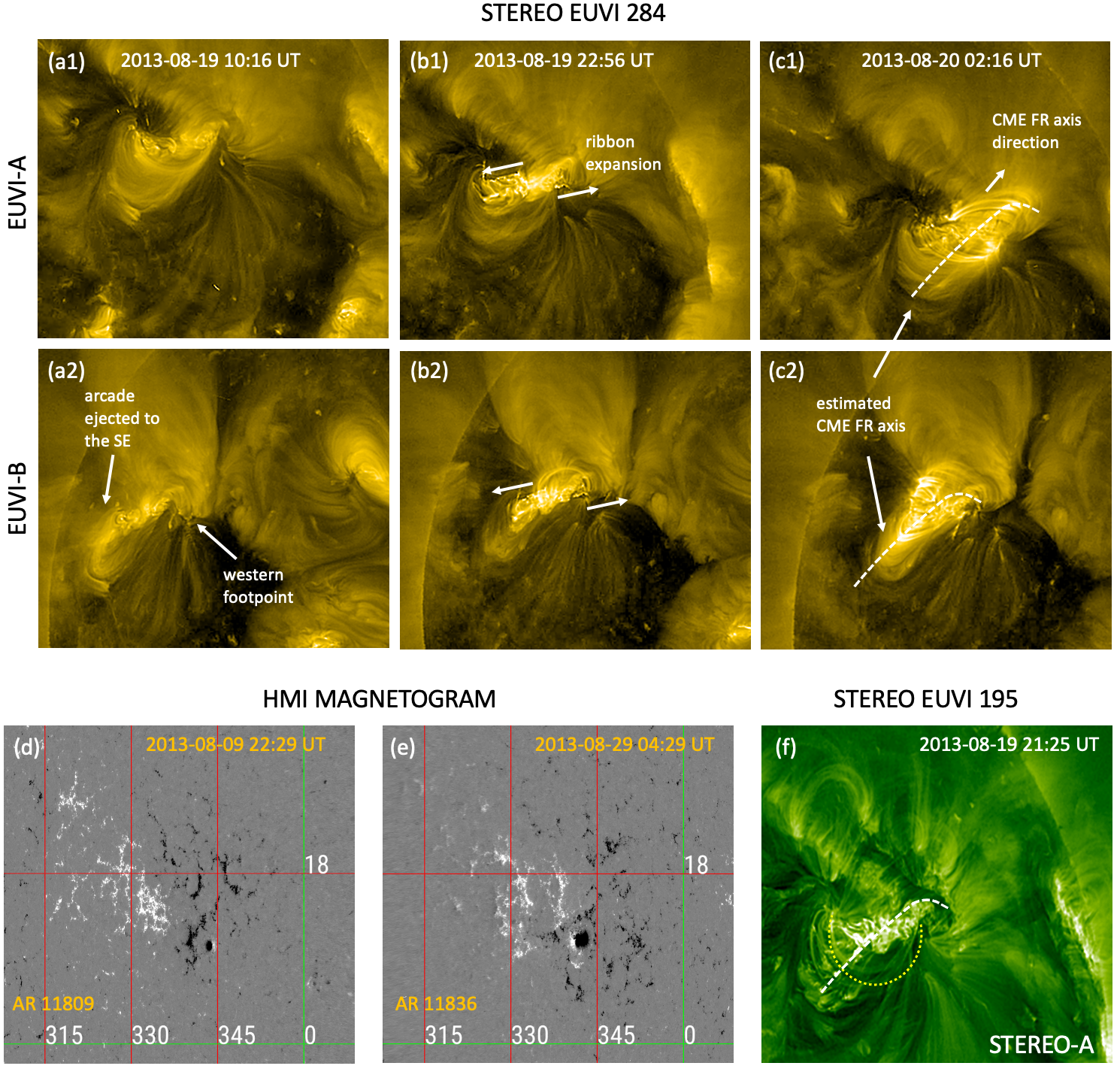}
\caption{Solar disk observations of the AR. \textit{Upper part:} Simultaneous snapshots of the source region in EUVI-A (top row) and EUVI-B (bottom row) 284 \AA\ channel images. (a) The AR loops exhibit a 'sigmoidal'-like pre-eruptive morphology. The approximate eruptive volume and CME footpoints are also shown. (b) The arrows point to the direction of ribbon expansion during the eruption, which indicates the likely orientation of the erupted MFR. (c) The dashed line representing our estimate for the shape of the MFR axis is compared to the overall shape of the PEAs. The full movie of the event is available online. \textit{Lower part}: (d) and (e) Evolution of the AR in magnetograms as observed by SDO/HMI, where positive (negative) polarity is indicated by white (black) colour. The images are taken from JHelioviewer using the latitudinal projection and Carrington coordinates. As the AR is located near the far-side central meridian from Earth's perspective, the magnetograms correspond to the times before the CME (d), when the AR is located close to the west limb, and after the CME (e), when the AR is located close to the east limb. The NOAA AR numbers are given on the legend. (f) EUVI-A 195 {\AA} image, where the PEA orientation is indicated with the white dashed curve and the coronal loops overlaying the pre-eruptive flux rope are indicated with the yellow dashed curve (details given in the main text). The wavelet enhanced EUVI-A and EUVI-B images are obtained after processing the original data with the technique described in \cite{Stenborg2008} and are available at \url{http://solar.jhuapl.edu/Data-Products/EUVI-Wavelets.php}. }
     \label{fig:euv_overview}
\end{figure*}

The CME erupted from the National Oceanic and Atmospheric Administration (NOAA) AR number 11809, located $177^\circ$ east from the central meridian as seen from Earth, in the northern solar hemisphere (E177N08). The AR had a relative simple $\alpha$ magnetic configuration when it was visible on the Earth-facing disk but it gave rise to several C-class flares. The region continued to evolve, with several extreme ultraviolet (EUV) brightenings during its passage across the STEREO-A-facing disk, including an apparent flux emergence episode on August 19 at $\sim$4:00~UT that led to significant outburst at $\sim$8:00 UT, which did not result in a CME. The AR loops exhibited a sigmoidal morphology in the EUVI-A hot channel (284 \AA), since  August 15, at least. Both EUVI-A and EUVI-B observed extended low-lying loops at the trailing part of the AR (Fig.~\ref{fig:euv_overview}a). All these observations indicate, albeit indirect due to the unavailability of photospheric magnetic field information, that shear was  building up withing AR 11809. The eruption sequence, as traced by the EUV erupting signatures and post-eruption arcades (PEAs), plays a central role in interpreting the in-situ signatures of the event later on, so we go through it in some detail.

\begin{figure}[htbp]
 \includegraphics[width=9cm]{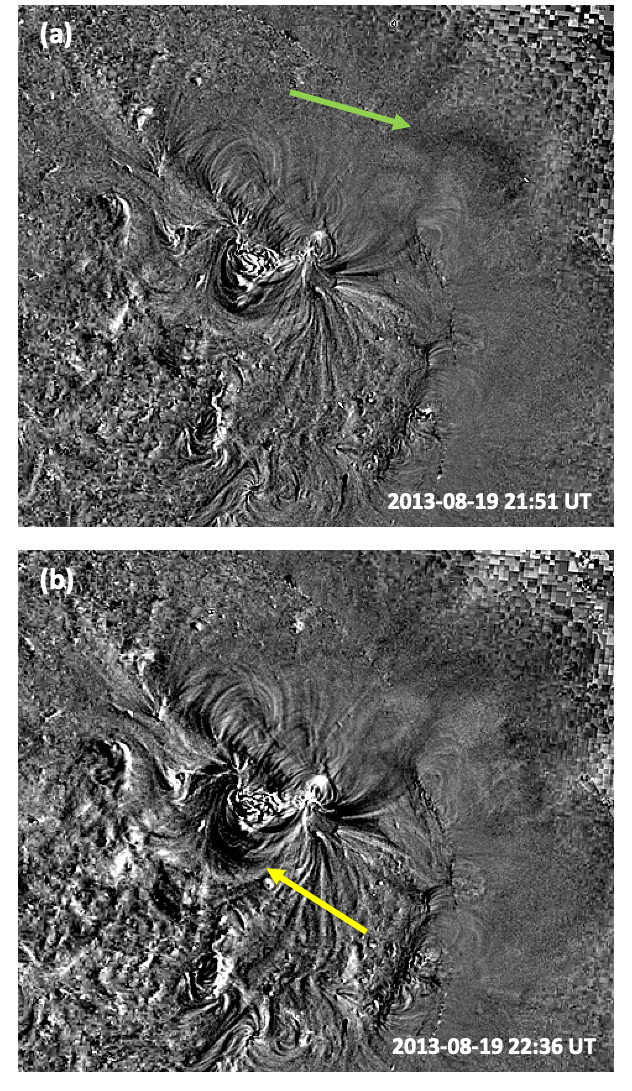}

\caption{EUVI-A 284 \AA\ channel base-difference images. The base image was taken at 21:31~UT on August 19. (a) A set of rising loops over the northwestern area of the source AR become evident by 21:51~UT and mark the STEREO-B-directed part of the CME. (b) A dimming along the southeastern part of the AR marks the side of the eruption towards STEREO-A. The full movie is available online. }
     \label{fig:euv_rundif}
\end{figure}

Starting at $\sim$21:51 UT on August 19, rising loops along the northwest side of the AR provide the first indication of the eruption (green arrow in Fig.~\ref{fig:euv_rundif}a). The eruption proceeds rapidly across the AR towards the southeast, as marked by the direction of the ribbon expansion (Fig.~\ref{fig:euv_overview}b) and an expanding dimming at that area by $\sim$22:36 UT (yellow arrow in Fig.~\ref{fig:euv_rundif}b). The highly non-radial orientation of the AR loops prior to the eruption, and the coronagraph signatures (discussed in Sect. \ref{sec:GCS_analysis}) indicate that the initial CME expansion was also highly non-radial, with the northwest part heading towards STEREO-B (roughly $40^\circ$ from the local AR radial), while the southeastern loops erupted eastwards of STEREO-A (i.e., beyond $40^\circ$ from the local AR radial). This expansion can be thought off as the spread of a 'folding hand fan'. Importantly, it occurs low in the corona, which explains the large width deduced from the coronagraph imaging, the forward modeling of the coronagraph observations and the in-situ observations. Such rapid expansion is also consistent with the appearance of a wide shock sheath in the coronagraph imagery. The CME-driven shock, roughly outlined with the yellow dashed curves in Fig.~\ref{fig:Orbit_solar_wind} a--c, was discussed in detail in Paper I and therefore is not discussed any further here. Relevant for this study is that the shock nose is oriented to W173N02, closer to MESSENGER and STEREO-A locations than to STEREO-B, and the width at 21.5 R\textsubscript{$\odot$} is 116$^{\circ}$. 

Another important clue for the CME magnetic configuration is provided by the orientation of the PEAs during the event. The ribbon expansion suggests a roughly east-west orientation (Fig.~\ref{fig:euv_overview}b) but the final PEAs appear to be more tilted (Fig.~\ref{fig:euv_overview}c). This time history suggests that the shape of the erupted MFR may deviate from the nominal semi-circular (`croissant'-like) shape. The MFR may have instead an undulating axis similar to the dashed line in Fig.~\ref{fig:euv_overview}c, possibly reflecting the sigmoidal appearance of the AR prior to the eruption. For the sake of brevity, we refer to this morphology as a `curved axis' in the rest of the paper. 
We posit that the PEAs act as a proxy for the orientation of the CME flux rope. Specifically, the view of the PEA evolution from the two EUVI 284 \AA\  viewpoints indicates that the western part of the CME may have a lower inclination relative to the solar equator than the trailing eastern part. As we will see later in Sect. \ref{sec:CME_flux_rope_type}, the PEA orientation helps to explain the MFR axis, as derived from the in-situ reconstructions. 

\begin{figure}[htbp] 
   \resizebox{\hsize}{!}{\includegraphics{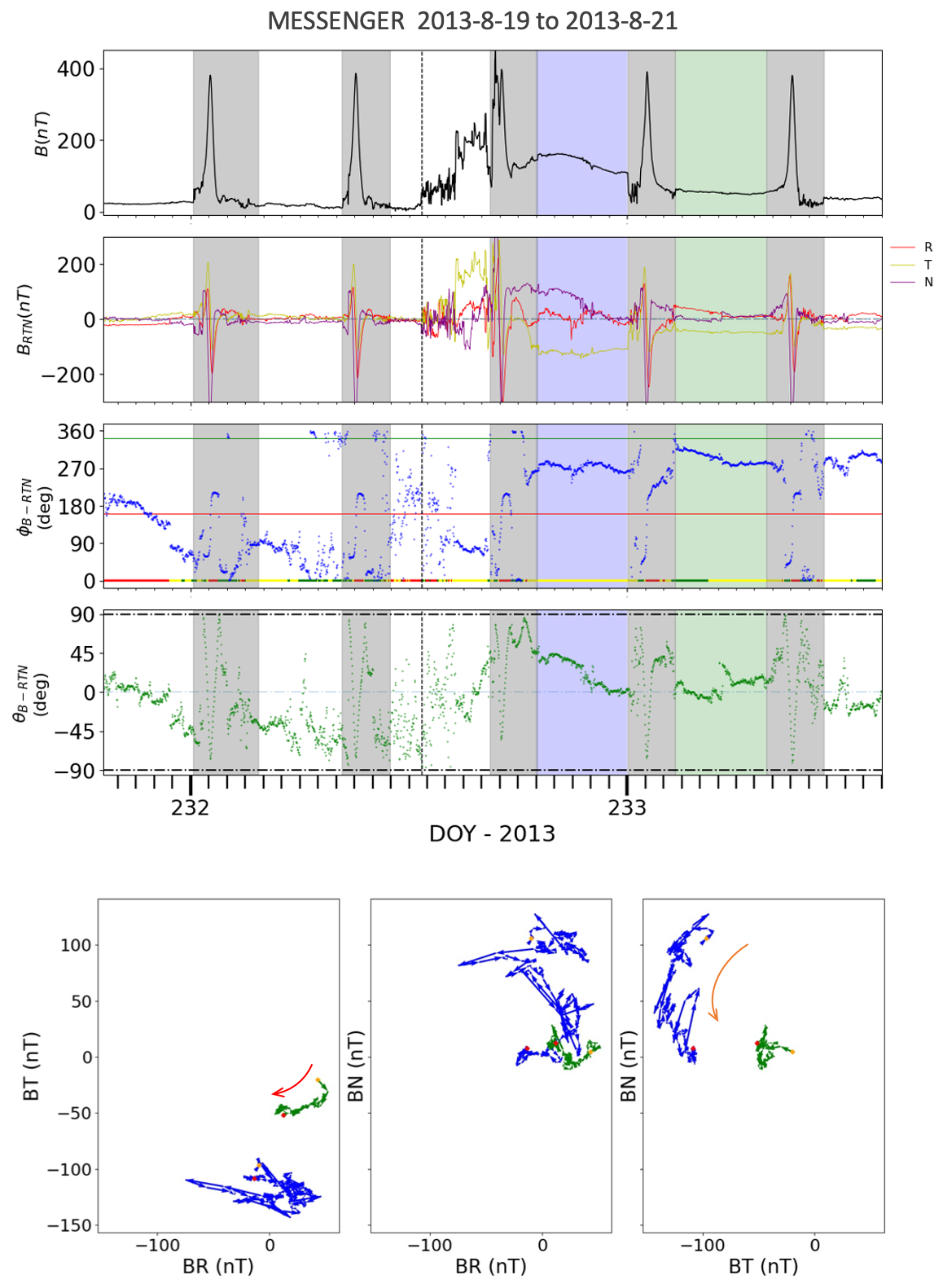}}
     \caption{Magnetic field observations by MESSENGER. \textit{Top}: The panels show \textit{from top to bottom}, the magnetic field magnitude, magnetic field components B\textsubscript{RTN}, and magnetic field azimuthal and latitudinal angles, $\phi$\textsubscript{B-RTN} and $\theta$\textsubscript{B-RTN}. The vertical dashed line represents the IP shock arrival and the color shaded areas mark the observed MC (blue) and MC-like structure (green). Red and green colour lines in $\phi$\textsubscript{B-RTN} angle panel (third) respectively denote inward and outward polarity, estimated from the magnetic field azimuth. The lower band shows the observed magnetic field polarity, where the yellow intervals represent the periods with the magnetic field oriented close to perpendicular to the nominal Parker spiral. The grey shaded areas delimit the intervals related to Mercury's magnetosphere, identified by eye. \textit{Bottom}: The panels show \textit{from left to right} B\textsubscript{T}-B\textsubscript{R}, B\textsubscript{N}-B\textsubscript{R} and B\textsubscript{N}-B\textsubscript{T} hodograms of the magnetic field for the periods indicated with the blue and green shaded areas in the top panels. The yellow and red points show the start and end of the hodogram for the selected periods, where the small arrows indicate the direction of the magnetic field. The orange (red) arrow indicates the direction of the rotation (partial rotation) observed within the MC (MC-like structure).}
     \label{fig:messenger}
\end{figure}
\begin{figure*}[htbp]
\centering
    \resizebox{\hsize}{!}{\includegraphics{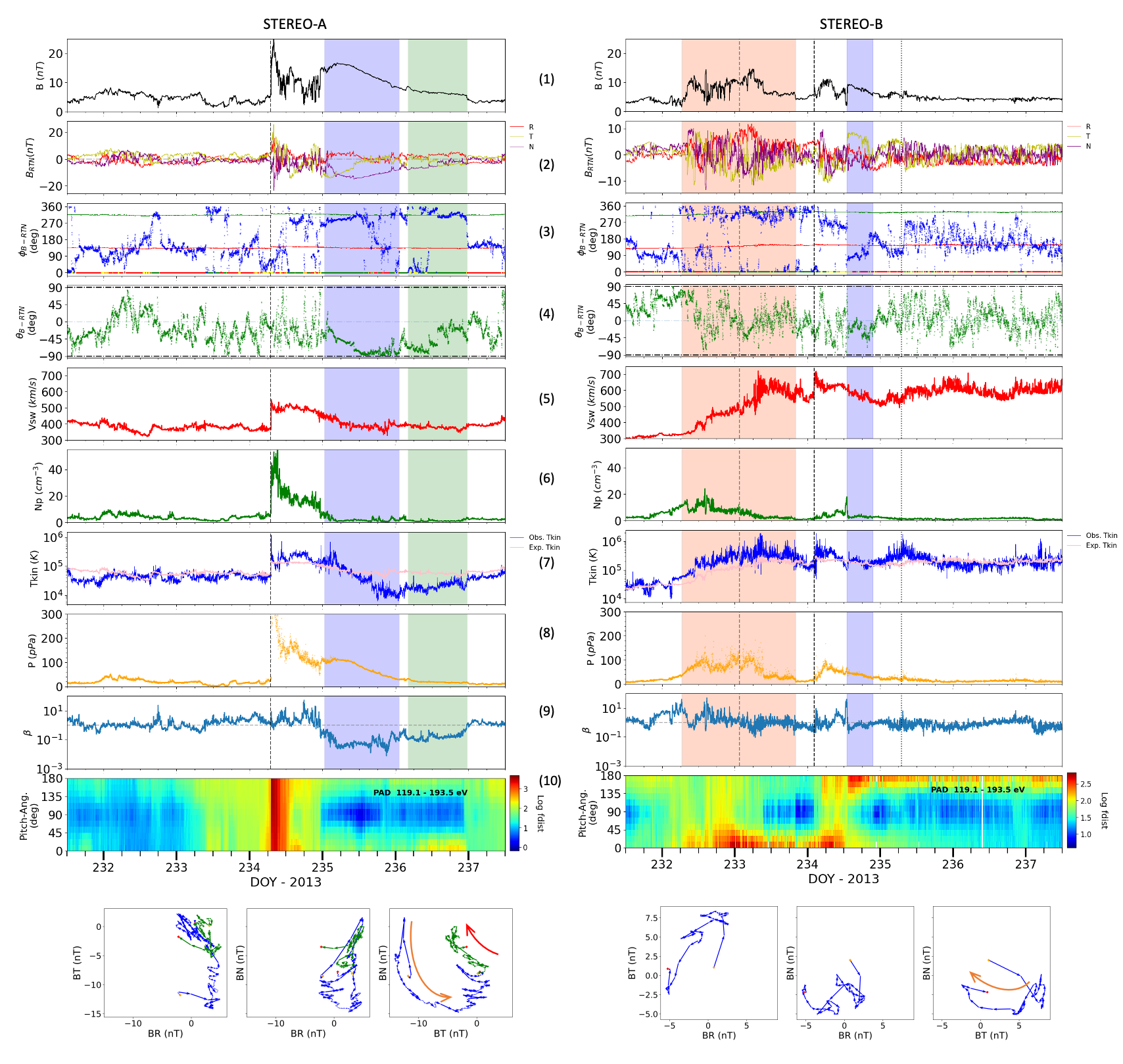}}
     \caption{In-situ plasma and magnetic field observations by STEREO-A (\textit{left}) and STEREO-B (\textit{right}). \textit{Top}: The panels show \textit{from top to bottom}, the magnetic field magnitude, the magnetic field B\textsubscript{RTN} components, the magnetic field azimuthal and latitudinal angles, $\phi$\textsubscript{B-RTN} and $\theta$\textsubscript{B-RTN}, the one-minute-averaged solar wind speed, the proton density, the observed and the expected proton temperature, the total pressure \citep[e.g.,][]{Russell2005}, where the total plasma pressure is estimated based on \cite{Mullan2006}, the plasma $\beta$, and the suprathermal (119 to 193 eV) electron intensity pitch-angle distribution, where the pitch angles are defined relative to the local magnetic field direction. The color lines and colour bands in panels (3) are the same as in Fig. \ref{fig:messenger}. The pink line shown in panels (7), corresponds to the expected kinetic temperature for ambient solar wind, calculated as explained in \cite{Elliott2012}. IP shock transits, MCs and MC-like structures are respectively indicated with black dashed lines and blue and green shaded areas. A SIR is marked by the salmon shaded area based on the referenced catalogue. \textit{Bottom}: The hodogram of the magnetic field components as in Fig. \ref{fig:messenger}.}
     \label{fig:icme_sta_stb}
\end{figure*}
\subsection{In-situ observations}
\label{sec:OBSER_in_situ}

To classify the different in-situ signatures within the ICME we consider the following criteria. We define the ICME start with the IP shock, followed by the sheath region and by the MO. Within the MO, the core of the structure, the MC is restricted to periods where the following features are shown: (1) an increase in the magnetic field strength, (2) a monotonic magnetic field rotation (flux rope) resulting in large net rotation of at least one of the magnetic field components, (3) low proton plasma temperature, and (4) plasma $\beta$ below 1 \citep{Burlaga1981}.
Thus, the MO includes more complex configurations of the CME core part, specifically, when more than one MFR or MC are identified, or if no monotonic rotation of the magnetic field direction is present within the MO. We note that throughout the paper we used the concepts of MC and MFR as measured in situ interchangeably to refer to the magnetic flux-rope signatures based on \cite{Burlaga1981} summarized above. 

\begin{table*}[htbp]
\centering
\caption{IP shock parameters obtained from the in-situ reconstruction from each spacecraft.}
\label{tableShockENLILFRONT}
\begin{tabularx}{0.8\textwidth}{cccccccccc} 
\hline
\hline
Spacecraft\ &r &Long&Lat&Shock arrival & Vshock &$\theta\textsubscript{Bn}$ & \multicolumn{1}{c}{V\textsubscript{d}/V\textsubscript{u}}&\multicolumn{1}{c}{B\textsubscript{d}/B\textsubscript{u}}&B\textsubscript{max}\\
& (au)&\multicolumn{2}{c}{(HGI, deg)}&(DOY in 2013)& (km s\textsuperscript{-1}) & (deg)&\multicolumn{2}{c}{(-)}&(nT)\\
 \hline
(1)&(2)&(3)&(4)&(5)&(6)&(7)&(8)&(9)&(10)\\
 \hline
MESSENGER &0.34 &61&1&232.53 &975  & 50  & - &3.3&224.0\\
STEREO-B &1.02 &116&-7&234.09 &615  &41 & 1.2 &1.4&11.0\\
STEREO-A &0.97 &38&-5&234.29  &483  & 84  &  1.4 &2.8&23.7 \\

 \hline
\end{tabularx}
\begin{flushleft}

 \footnotesize{ \textbf{Notes.} Column 1: Observing spacecraft. Columns 2, 3 and 4: Radial distance, longitude and latitude of spacecraft in heliographic inertial coordinates \citep[HGI,][]{Burlaga1984} at the time of shock arrival to STEREO-A (used as a reference). Columns 5 to 7: Observed shock arrival time, shock speed (shock transit speed for MESSENGER), and angle between the shock normal and the upstream field. Columns 8 and 9: Ratio between downstream (d) and upstream (u) solar wind speed and magnetic field strength. Column 10: Maximum magnetic field strength observed within the ICME. 
 }

\end{flushleft}
\end{table*}

The magnetic field observations by MESSENGER are shown in Fig. \ref{fig:messenger}. The increase in the magnetic field magnitude at DOY 232.53 UT marks the arrival of the IP shock (vertical dashed line). The shock arrival coincides with an abrupt decrease in SEP intensities, namely, near-relativistic electrons (Fig. 12 in Paper I). There are no bi-directional suprathermal electrons (BDE) nor plasma data available. Thus, the MO identification is only based on visual inspection of the magnetic field. The MO starts at DOY 232.79, with a change in the magnetic field polarity (third panel) along with the presence of coherent and ordered magnetic field. Specifically, we observe a smooth and monotonic change on the magnetic field components. Based on the magnetic field profile, the fluctuation at the front of the structure may indicate magnetic reconnection processes associated with erosion \citep{2012Ruffenach,2017Kilpua}. The maximum magnetic field strength within the MO is $\sim$170 nT, which is the highest value registered by \cite{2015Winslow} study. They sampled 61 ICMEs observed by MESSENGER between 2011 and 2014, where the second highest value corresponds to $\sim$109 nT. The MO ends at DOY 233.32, when no more rotation is observed on the magnetic field components (second, third and fourth panel) and the magnetic field strength (first panel) is back to ambient solar wind level. 

The MO at MESSENGER consists of two structures. The first core structure, observed from DOY 232.79 to DOY 233.00 (blue shaded area in top panels of Fig. \ref{fig:messenger}), shows a southward rotation (MC) on the magnetic field components, as seen in the B\textsubscript{N}-B\textsubscript{T} hodogram (orange arrow in bottom panel of Fig. \ref{fig:messenger}). The second part of the structure, observed from DOY 233.11 to DOY 233.32 (green shaded area), leads us to different interpretations. The structure may be understood as the lagging part of the CME `tongue' \citep[][and references therein]{Cocconi1958,McComas1995}. Even so, the partial rotation (red arrow in bottom panel of Fig. \ref{fig:messenger}) in the magnetic field direction displays signatures of a MC-like structure \citep[F- for][]{Nieves-Chinchilla2019}. Thus, the MO as observed by MESSENGER, might be classified as `complex' \citep{Lepping2006} due to the two configurations present in the magnetic structure.

In-situ plasma and magnetic field observations by STEREO are shown in Fig. \ref{fig:icme_sta_stb}. Left panels show that the IP shock (black dashed line) arrives at STEREO-A at DOY 234.29, represented as an abrupt increase in magnetic field strength (1), proton plasma speed (5), density (6), and total pressure (8). The MO lasts from DOY 234.97 to 236.98, observed as an increase of the magnetic field strength (1), a depression in the proton plasma density (6), a gradual decrease in the proton speed (5) along with a decrease in the plasma $\beta$ (9), and lower-than-expected temperature (7), due to the expansion present in the MO. The ICME was bounded by low-speed solar wind streams and there is a sector boundary crossing at the beginning of DOY 231 (not shown). STEREO-A observes a sudden decrease in SEP intensities, namely, protons up to 30 MeV, at the time of the IP shock arrival, and also a transient depression $\sim$12 hours later (Fig. 2 in Paper I), probably related to the closed magnetic field topology of the MO.

The MO at STEREO-A is compounded by two different structures, one MC (blue shading in left top panels of Fig. \ref{fig:icme_sta_stb}) and one MC-like (green shading). The MC-like structure is observed just after a change in some of the parameter profiles, at the beginning of DOY 236. We observe an increase in temperature (7) and plasma $\beta$ (9), and a change in the apparent polarity from negative to positive inferred from the local magnetic field vector (3). As presented in the bottom hodogram, the first MC is observed as a southward rotation (orange arrow in left bottom panel of Fig. \ref{fig:icme_sta_stb}) from DOY 235.02 to 236.05 (blue shaded area), and the second MC-like structure as a partial northward rotation (red arrow) from DOY 236.17 to 236.98 (green shaded area). The first MC seems to be dominant within the MO. The expansion is clearly observed in the first MC, while in the second MC-like structure the solar wind speed is constant (left panel 5 in Fig. \ref{fig:icme_sta_stb}). Thus, the MO as observed by STEREO-A might also be classified as complex due to the two configurations present in the magnetic structure. 

Right panels of Fig. \ref{fig:icme_sta_stb} show solar wind plasma and magnetic field observations by STEREO-B. The ICME was bounded by high speed solar wind streams. A stream interaction region (SIR, salmon shaded area) is observed just before the shock arrival. The IP shock arrives at STEREO-B at DOY 234.09 (black dashed line), with a moderate increase in magnetic field strength (1) and solar wind speed (5). The SEP profile at STEREO-B is not influenced by the IP structure arrival (Fig. 2 in Paper I). Right after the shock, signatures of BDE are observed (10), with an increase in the intensity of the pitch-angle distribution at 0{$^{\circ}$} and 180{$^{\circ}$}, which might be associated only with the sheath, as the different parameters do not show clear MO signatures, such as low plasma $\beta$ (9). However, around DOY 234.54 there are signatures suggesting that STEREO-B is crossing a MO edge and possibly also the heliospheric current sheet (HCS), as a discontinuity in the magnetic field magnitude (1) and plasma density (6) and an apparent change in the magnetic field polarity (3) are observed. We note that the sector boundary crossing is different from the one observed at STEREO-A. From DOY 234.54 and during $\sim$8 hours there are signatures of a MFR topology. A eastward rotation (MC) is present in the B\textsubscript{T}-B\textsubscript{N} hodogram (bottom, orange arrow), coinciding with a period of increased magnetic field magnitude (1), solar wind speed decay (5), decreased density (6), and low plasma $\beta$ (9). At 234.89, the spacecraft seems to be back in the sheath region, that lasts until the end of the ICME at DOY 235.29. Alternatively, the possible interaction with the HCS could have eroded the rear part of the MFR and therefore not showing clear signatures of magnetic field rotation. The magnetic field reconnection could also be related to the not very low plasma $\beta$ values (9) measured inside the MO, which are very close to 1  \citep{2012Ruffenach}. The ICME ending is indicated with the vertical dotted line, when a sharp change of magnetic field direction (3) is observed and the total pressure (8) returns to ambient level, suggesting the presence of a boundary.

A summary of the IP shock parameters obtained from in-situ reconstruction is shown in Table \ref{tableShockENLILFRONT}. In case of STEREO spacecraft, the shock speed, the angle between the shock normal and the upstream field ($\theta$\textsubscript{Bn}), and the ratios of downstream to upstream solar wind speed (V\textsubscript{d}/V\textsubscript{u}) and magnetic field intensity (B\textsubscript{d}/B\textsubscript{u}), are taken from the IP shock catalogue, maintained by the University of Helsinki\textsuperscript{\ref{footnote IP Helsinki}}. The methods used for estimating each of the IP shock parameters are indicated in the catalogue documentation. Column 6 shows that STEREO-B is observing a higher shock speed than STEREO-A. Column 7 indicates that STEREO-A observes a quasi-perpendicular shock ($\theta\textsubscript{Bn}$=84$^{\circ}$) while STEREO-B might observe an oblique configuration ($\theta\textsubscript{Bn}$=41$^{\circ}$). We note that the shock normal angle and B\textsubscript{d}/B\textsubscript{u} are mostly consistent with the values in Level 3 shock list of STEREO\textsuperscript{\ref{footnote stereo_level3}}. The shock parameters presented for MESSENGER
are based on the magnetic coplanarity method \citep{1966Colburn,1971Lepping}.
Column 7 shows that $\theta\textsubscript{Bn}$ is 50$^{\circ}$. Thus, the shock configuration is oblique at MESSENGER. 

\section{CME analysis} \label{sec:Remote-sensing analysis}
In this section, we present the 3D reconstruction of the CME from $\sim$3.5 to $\sim$13 R\textsubscript{$\odot$}, taking advantage of the multi-point view from STEREO-A, STEREO-B and SOHO. We also include the MFR type derived from the solar disk observations. 
\subsection{CME reconstruction}
\label{sec:GCS_analysis}
 \begin{figure*}[htbp] 
\centering
\includegraphics[width=12cm]{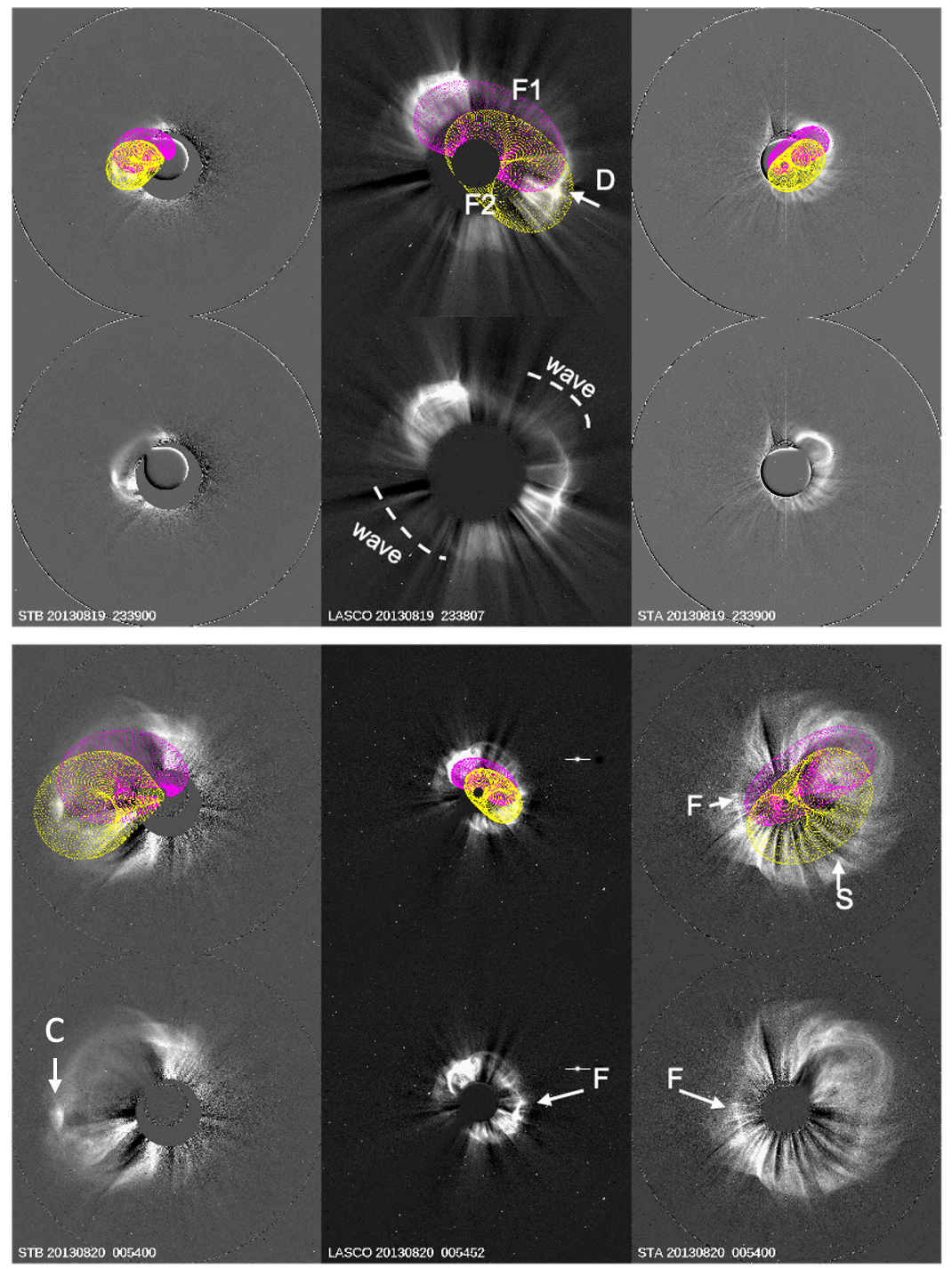}
\caption{3D reconstruction of the CME MFR. Only the fits at two instances are shown. A single MFR is fitted with two GCS models (F1 and F2) to account for the curvature of the MFR axis, as indicated by the EUV observations. The extent of the CME-driven shock (or wave) is indicated at the top panel. `D' points to a dimple at the MFR front and `F' to the development of a flank at the north of the dimple. `S' points to an area where the F1-F2 GCS fits fails. Pileup feature `C' indicates a distorted region of the MFR. These features are discussed in detail in the text.}
     \label{fig:GCS_recon}
\end{figure*}
 
 We performed the 3D reconstruction of the CME using the scraytrace forward modeling framework available in the SECCHI IDL SolarSoft distribution\footnote{\url{http://www.lmsal.com/solarsoft/}}. We fitted the CME MFR with the widely-used GCS model \citep{Thernisien2006GCS,Thernisien2011}. We adopted, however, a different approach than in Paper~I. In that work, we were primarily interested in the properties of the CME-driven shock and the overall CME kinematics for making the connection to the energetic particle measurements. In this study, we are interested in a detailed comparison of the magnetic structure of the CME as derived by remote observations and in-situ measurements. Therefore, we pay a lot of more attention on the shape of the CME MFR derived from the GCS fit and its consistency with the other remote observations. 
 
 A close inspection of Fig.~11 in Paper~I indicates that the GCS fit was unable to account fully for the CME front in all three views, particularly for the COR2-A images. This discrepancy along with the PEA evolution, discussed in Sect. \ref{sec:OBSER_remote-sensing}, led us to suspect that the CME MFR is likely to have a curved axis, which, however is a constraint of the GCS model. To circumvent this constraint and to check whether a curved axis will make a better fit to the multi-viewpoint white light signatures, we fitted two slightly different GCS models to the observed CME (Fig.~\ref{fig:GCS_recon}). We refer to them as F1 (magenta) and F2 (yellow). We want to make it clear that we are not considering two separate CMEs or even two separate phases of a single eruption. Instead, we propose that the actual CME MFR is the convolution of the two GCS models in Fig.~\ref{fig:GCS_recon}; namely, it may be an MFR with an increasing width along the west-to-east direction, as well as, with a slightly  curved axis.

 We performed the GCS fit for six snapshots while the CME was within the COR2 and LASCO field of views. The main parameters of the fits are shown in Table~\ref{table:GCS_fit}. The half-widths of F1 and F2 along the ecliptic plane were estimated based on \cite{Dumbovic2019}, where the half angular width in the equatorial plane is represented by ${R\textsubscript{maj}-{(R\textsubscript{maj}-R\textsubscript{min})} \times |tilt|/90}$. It results in 70$^{\circ}$ and 39$^{\circ}$ for F1 and F2 half angular width, respectively. We note that the width deviation based on the mean half-angle error \citep{Thernisien2009} is estimated as +13$^{\circ}$/-7$^{\circ}$. 
 
 The F1 direction (W170) is closer to MESSENGER (W159) and STEREO-A (W144) than to STEREO-B (E138 or W222). However, the F1 fit is wide ($140^\circ$) and hence the CME western flank extends to W240. Thus, the CME flank is likely to intercept STEREO-B, as measured in situ (Sect. \ref{sec:OBSER_in_situ}). The F2 portion of the CME is directed even closer (W160) toward MESSENGER and STEREO-A, with the eastern flank extending to W121. Thus, the F1 and F2 fits indicate that the CME MFR angular extent in the equatorial plane is 119$^{\circ}$and it is propagating self-similarly from 23:54 UT on August 19 until the end of the reconstructed period at 00:54 UT on August 20. The large width explains why the CME was intercepted by all three spacecraft. Finally, the differences in the F1 and F2 reconstructions, such as the ratios (Table~\ref{table:GCS_fit}) and speeds (the CME speed is $\sim$969 km s\textsuperscript{-1} along the F1 portion, and $\sim$1237 km s\textsuperscript{-1} along F2) indicate asymmetries across the CME structure that are likely reflected in the in-situ measurements, as we discuss later. 
 
 Regarding the CME MFR orientation, the F1-F2 tilts indicate that a curved axis, varying from $22^\circ$ on the west to $36^\circ$ on the east offers a much better fit to the CME white light signatures. This orientation is consistent with the low corona EUV signatures, particularly the PEAs. There are, nevertheless, a couple of locations where even the double F1-F2 fit is inadequate. These locations warrant some consideration and are marked with the letters 'C', 'D', `F' and `S' in Figure~\ref{fig:GCS_recon}. The discrepancy at location `D' is rather straightforward to understand. The MFR appears to propagate along the position angle of a streamer stalk where increased plasma pressure often distorts CME fronts \citep[e.g.,][]{manchester_iv_coronal_2005,das_evolution_2011}. This dimple is apparent only from the LASCO viewpoint, which suggests that the direction of propagation is between LASCO and STEREO-A. The northern flank of the dimpled front, eventually appears in the COR2-A images (point F). A similar feature, labeled C, appears in the COR2-B images, indicating that the F2 front may be strongly distorted along the position angle. An inspection of the three snapshots at the bottom row of Figure~\ref{fig:GCS_recon} suggests that features C and F may be related and they presumably mark the two flanks of an extended distortion/pileup across the CME. The implications of these features is that STEREO-A and MESSENGER are likely to encounter a distorted/compressed CME front since the F2 part of the CME is propagating towards the locations of these spacecraft. Finally, the F2 fit is also unable to fully match the CME front along the south (feature S). Considering the disagreement in F, we conclude that the CME MFR is even wider along its eastern and southern parts than we are able to accommodate with a single GCS model. This finding points further to our EUV-based interpretation that the MFR reflects the large width of the EUV loops along southeastern part of the AR in Fig. \ref{fig:euv_rundif}b. 
 
 In summary, the GCS fits suggest that a wide and slightly curved MFR is expected to be intercepted by STEREO-B, MESSENGER, and STEREO-A, with direct hits only at MESSENGER and STEREO-A. This configuration is represented in Fig. \ref{fig:Orbit_solar_wind}, right. The orange arrow indicates the averaged CME apex direction and the orange lines outline the estimated width of the CME. The curvature of the MFR axis further suggests that the in-situ measurements at MESSENGER and STEREO-A, and STEREO-B may exhibit different signatures. We discuss these measurements in Sect. \ref{sec:in situ analysis}.

\begin{table*}[htbp]

\caption{3D CME properties derived from the GCS fits shown in Figure~\ref{fig:GCS_recon}.}
\label{table:GCS_fit}.
\begin{tabularx}{0.95\textwidth}{ccccccccccccccccc} 
 
\hline
\hline
 Date-Time & \multicolumn{2}{c}{Lon}&\multicolumn{2}{c}{Lat}&\multicolumn{2}{c}{Tilt}&\multicolumn{2}{c}{Height}&\multicolumn{2}{c}{Half- angle}&\multicolumn{2}{c}{Ratio}&\multicolumn{2}{c}{$R\textsubscript{maj}$} &\multicolumn{2}{c}{$R\textsubscript{min}$}\\
 (UT in 2013)& \multicolumn{2}{c}{(deg)} & \multicolumn{2}{c}{(deg)} & \multicolumn{2}{c}{(deg)} & \multicolumn{2}{c}{(R\textsubscript{$\odot$})} & \multicolumn{2}{c}{(-)}& \multicolumn{2}{c}{(deg)}& \multicolumn{2}{c}{(deg)}& \multicolumn{2}{c}{(deg)}\\
   & F1 & F2 & F1 & F2 & F1 & F2 & F1 & F2 & F1 & F2&F1&F2&F1&F2&F1&F2\\
\hline
(1)&(2)&(3)&(4)&(5)&(6)&(7)&(8)&(9)&(10)&(11)&(12)&(13)&(14)&(15)&(16)&(17)\\
 \hline

 \hline
08-19 23:24 & 172 & 160 & 13 & -10 & 22 & 36 &  5.2 &  5.0 & 50 & 25 & 0.35&0.40&70&49&20&24 \\
08-19 23:39 & 171 & 160 & 13 & -10 & 22 & 36 &  6.3 &  6.9 & 52 & 25& 0.35&0.40&72&49&20&24 \\
08-19 23:54 & 171 & 160 & 13 & -10 & 22 & 36 &  7.1 &  8.2 & 65 & 25 & 0.35&0.40&85&49&20&24\\
08-20 00:24 & 171 & 160 & 13 & -10 & 22 & 36 &  9.8 & 11.1 & 65 & 25& 0.35&0.40 &85&49&20&24\\  
08-20 00:39 & 170 & 160 & 13 & -10 & 22 & 36 & 11.4 & 13.2 & 65 & 25& 0.35&0.40&85&49&20&24\\
08-20 00:54 & 170 & 160 & 13 & -10 & 22 & 36 & 12.6 & 14.7 & 65 & 25& 0.35&0.40&85&49&20&24\\
\hline
\\
\end{tabularx}

\footnotesize{ \textbf{Notes.} Column 1: Date and time UT in 2013. Columns 2 to 5: Stonyhurst coordinates of the F1 and F2 leading-edge (LE) orientation. Columns 6 to 13: F1 and F2 angle with respect to the solar equator, height from the Sun center, half-angle and aspect ratio. Columns 14 to 17: F1 and F2 face-on half-width ($R\textsubscript{min}$ + half-angle), and edge-on half-width ($\arcsin(ratio)$) according to \cite{Thernisien2011}. 
}
\end{table*}

\subsection{CME MFR type}
\label{sec:CME_flux_rope_type}
The intrinsic MFR type of the CME can be determined from the solar disk observations and this information can be compared to the in-situ measurements. We follow the methodology used by \cite{Palmerio2017}, who described how the magnetic helicity sign (chirality), the MFR axis orientation (tilt), and the axial magnetic field direction can be estimated from several known indirect proxies on the solar disk. For the event analyzed here, some of the proxies that rely on photospheric magnetogram information cannot be used because the source region is located on the far-side.

Figures \ref{fig:euv_overview}d and e show the photospheric evolution of AR as observed in SDO/HMI line-of-sight magnetograms two weeks prior -NOAA AR 11809- (d) and two weeks after -NOAA AR 11836- (e) the CME. Clearly, the AR has evolved during its rotation. The leading spot has grown into two spots and there is a mixture of positive and negative polarities around the sunspots. These are indications of flux emergence during the AR far-side passage. This is in agreement with the EUV activity prior to the August 19 CME, as we discussed earlier, and demonstrates the value of EUV far-side imaging for understanding the AR evolution.

Based on \cite{Palmerio2017}, the orientation of the MFR axis can be considered approximately parallel to the polarity inversion line \citep[PIL,][]{Marubashi2015}, or to the orientation of the PEAs, associated with the reconnection that occurs behind the CME \citep{Yurchyshyn2008}. For this eruption, it is not possible to estimate the orientation of the PIL from the  magnetograms but we can use the ribbons as a proxy (Figure~\ref{fig:euv_overview}b). They present a tilt of $\sim$14$^{\circ}$, which is somewhat less that the tilt of the western part (F1) of the CME ($22^\circ$, Col. 6 in Table~\ref{table:GCS_fit}). The ribbon orientation, however, does not capture any additional shifts in the orientation of the coronal structures due to shear. The PEAs are a better proxy for the coronal MFR and are often the clearest signature of the CME eruption in the low corona visible in EUVI images \citep{TripathiBothmerCremades2004}. Fig. \ref{fig:euv_overview}c shows STEREO EUVI 284 {\AA} images, where the white dashed curves represent the PEA orientation, as observed by STEREO-A (c1) and STEREO-B (c2) associated with the CME eruption. Thus, the orientation of the MFR axis or tilt angle, |$\tau$|, seems to be $\sim$22$^{\circ}$ ($\sim$36$^{\circ}$) in the western (eastern) part of the MFR, as this is the angle the PEA line with respect to the heliographic equator, determined by eye. This is in very good agreement with the MFR tilt derived from the GCS model ($22^\circ$ and $36^\circ$, Cols. 6 and 7 in Table~\ref{table:GCS_fit}). Following the classification by \cite{Kilpua2011}, this value of the tilt angle corresponds to bipolar (parallel) flux ropes in situ, as |$\tau$| $\leq$ 45$^{\circ}$. We note that the mean PEA orientation uncertainty is ±10$^{\circ}$ \citep{Yurchyshyn2008}. 

In this particular eruption, the chirality of the flux rope of the CME could  be estimated using three of the six methods explained in \cite{Palmerio2017}. Firstly, we used the skew (acute angle) that the coronal loops overlying the pre-eruptive flux rope form with the PEA, if we consider it parallel to the tilt orientation. Figure \ref{fig:euv_overview}f shows that the pre-eruptive arcades, outlined with the yellow dotted curve, are right-skewed with respect to the white curve representing the PEA, when viewed from the positive polarity side (situated to the north of the line; polarity as in Fig. \ref{fig:euv_overview}d and e). As explained in \cite{Martin1998}, right-skewed with respect to the PEA implies positive magnetic helicity. Secondly, we used the orientation and displacement of the ribbons along the PIL (Fig. \ref{fig:euv_overview}b). If the PIL is vertical on the solar image, as the left ribbon is displaced downwards and the right ribbon upwards, this represents positive chirality \citep{Demoulin1996}. Finally, we used the hemispheric helicity rule, based on the tendency for the magnetic structures of the Sun to have negative (positive) helicity in the northern (southern) hemisphere \citep{Pevtsov2003}. Due to the location of the AR (N08), the chirality is expected to be negative. However, this method cannot be used as a reliable proxy as it is only true in 60-75\% of the ARs. Thus, we consider the chirality to be positive, as given by the skew of the coronal loops and the orientation and displacements of the ribbons. 

Lastly, the axial field orientation of the flux rope, can be taken as the direction of the magnetic field that runs nearly parallel to the PEA and that depends on the helicity of the source region. This direction could be inferred from Fig. \ref{fig:euv_overview}c and d. For a positive chirality, the field is directed left as seen from the positive polarity of the AR \citep{BothmerSchwenn1994}. This would mean northwest axial field direction, indicated in Fig. \ref{fig:euv_overview}c1. 

Thus, the prediction of the MFR type using solar disk observations, derives a positive chirality, a MFR tilt of $\sim$36$^{\circ}$ ($\sim$22$^{\circ}$) in the eastern (western) part and an axial field direction to the west. This corresponds to a low-tilted cloud type South-West-North (SWN) \citep[e.g.,][]{BothmerSchwenn1998}.

\section{ICME analysis}
\label{sec:in situ analysis}
In this section, we derived the heliospheric evolution of the ICME from the plasma and magnetic field data measured at MESSENGER, STEREO-B and STEREO-A locations, representing three different vantage points radially and longitudinally distributed in the heliosphere. We examined the in-situ data to estimate the global shape and geometry of the ICME, including the global IP shock surface, and the internal magnetic structure as observed by the different spacecraft. We used the EC model for the reconstruction of the MCs. 

\subsection{IP shock analysis}\label{sec:Analysis_IP_shock}

\begin{figure}[htbp] 
 \centering
  \resizebox{\hsize}{!}{\includegraphics{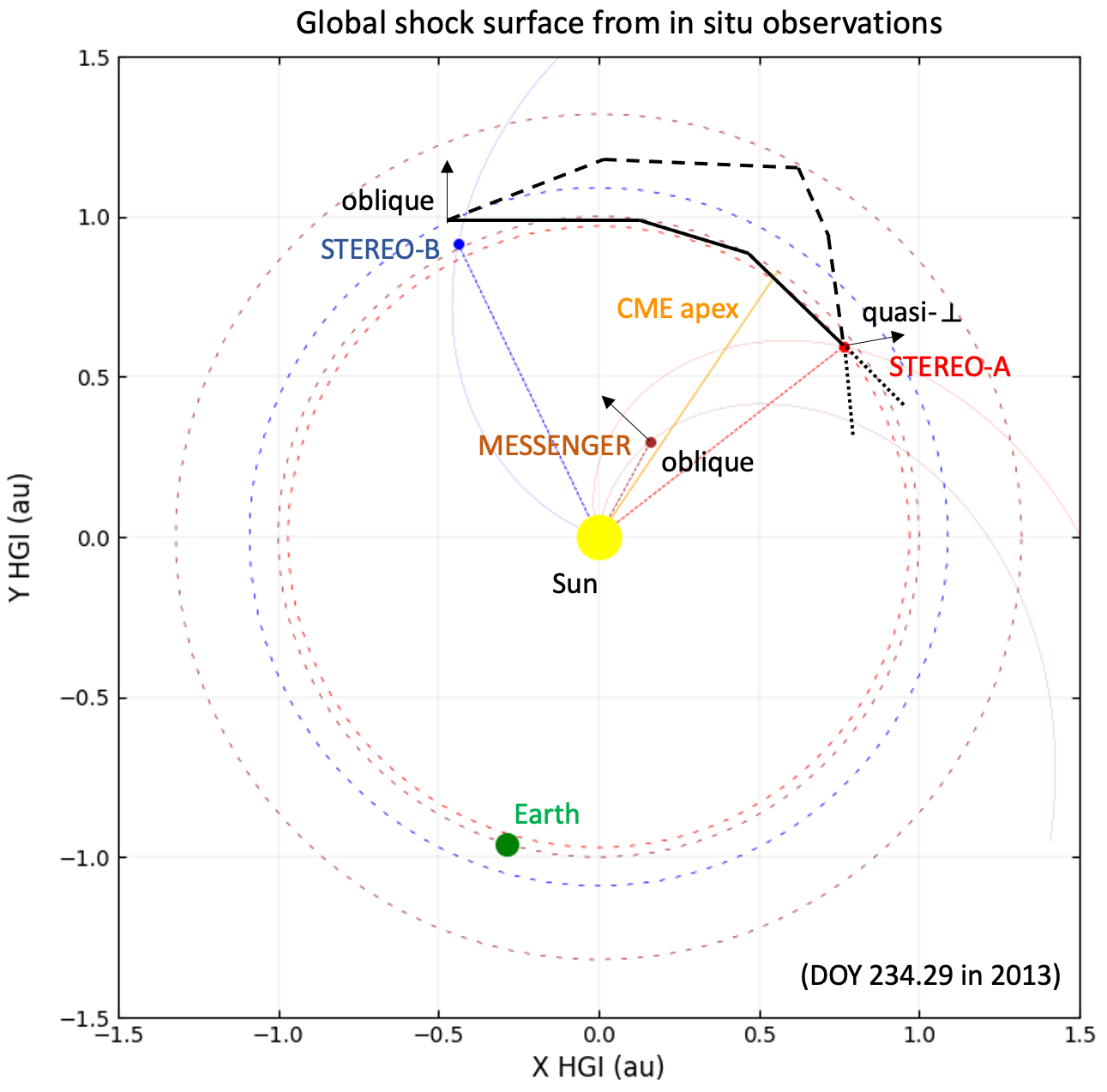}}
    \caption{Global configuration of the IP shock in the equatorial plane estimated from in-situ observations. The approximate shape of the IP shock is indicated with the black bold straight and dashed lines (details given in the main text). The arrows represent the shock normals and the configuration of the shock is indicated as oblique or quasi-perpendicular. The Parker spirals connecting each spacecraft are calculated using the in-situ solar wind speed observed at the locations of STEREO-A (also used for MESSENGER) and STEREO-B at the time given on the plot legend. The Earth (not to scale) is given for reference and the Sun at the center is indicated with the yellow circle (not to scale).}
    \label{fig:shock_anal}
\end{figure}


We analysed the shock parameters observed at the locations of MESSENGER, STEREO-B and STEREO-A, to relate the findings to the global shape of the shock. To simplify the analysis, we neglected the small separation angles in latitude between the spacecraft (Col. 4 in Table \ref{tableShockENLILFRONT}). As reference for the shock analysis, we considered the time of arrival of the IP shock at STEREO-A, t\textsubscript{S\_STA}=DOY 234.29 UT in 2013, and the radial distance of STEREO-A, r\textsubscript{STA}=0.97 au.  

As discussed in Sect. \ref{sec:OBSER_in_situ} and summarized in Col. 5 of Table \ref{tableShockENLILFRONT}, MESSENGER is the first spacecraft that observes the IP shock. Based on the narrow upstream and downstream regions selected for the shock fitting, the magnetic field compression ratio B\textsubscript{d}/B\textsubscript{u} is 3.3 (Col. 9 in Table \ref{tableShockENLILFRONT}), which represents a moderately strong shock \citep{2013Neugebauer}. Although STEREO-B is located 0.05 au further from the Sun than STEREO-A, STEREO-B observes the IP shock around five hours earlier than STEREO-A, and based on V\textsubscript{d}/V\textsubscript{u} = 1.2 (Col. 8) and B\textsubscript{d}/B\textsubscript{u} = 1.4 values, it encounters a weak shock. Finally, the V\textsubscript{d}/V\textsubscript{u} = 1.4 and B\textsubscript{d}/B\textsubscript{u} = 2.8 ratios represent that STEREO-A observes a moderately strong shock.  

\begin{table*}[ht]
\begin{minipage}{\textwidth} 
\caption{ICME signatures observed at different spacecraft.}
\label{table:ICME signtures}
\begin{tabularx}{1\textwidth}{cccccccccccccc} 
\hline
\hline
 & \multicolumn{2}{c}{ICME} & \multicolumn{2}{c}{MO}&\multicolumn{2}{c}{MC}&\multicolumn{2}{c}{MC-like}&V\textsubscript{sw}&B\textsubscript{max}&B\textsuperscript{'}\textsubscript{max}&V\textsubscript{sw}&FR\\
s/c&T\textsubscript{start}&T\textsubscript{end}&T\textsubscript{start}&T\textsubscript{end}& T\textsubscript{start}&T\textsubscript{end}&T\textsubscript{start}&T\textsubscript{end}&{ICME}&\multicolumn{3}{c}{MO} &type\\
&\multicolumn{8}{c}{(DOY in 2013)}&(km/s)&\multicolumn{2}{c}{(nT)}&(km/s)\\
\hline
(1)&(2)&(3)&(4)&(5)&(6)&(7)&(8)&(9)&(10)&(11)&(12)&(13)&(14)\\
 \hline

 \hline
MESS & 232.53 &233.32 & 232.79 &233.32& 232.79 &233.00& 233.11 & 233.32 &800\textsuperscript{a}&168.9&28.8&575\textsuperscript{a}&NES\\
STB & 234.09 &235.29 &234.54 & 234.89  & 234.54 & 234.75 &-&-&600&9.2&10.0&568&WSE \\
STA & 234.29 &236.98&  234.97 &236.98& 235.02  &236.05& 236.17 &236.98&455&16.6 &16.6&398&ESW\\

\hline
\\
\end{tabularx}
\footnotesize{ \textbf{Notes.} Column 1: Observing spacecraft. Columns 2 to 9: Start and end of the ICME, MO, MC and MC-like structures, respectively. Column 10: Mean solar wind speed within the ICME. Columns 11 to 13: Maximum magnetic field strength, scaled magnetic field strength to the heliocentric distance of STEREO-A (details given in the main text), and mean solar wind speed within the MO, respectively. Column 14: Type of field rotation \cite[e.g.,][]{BothmerSchwenn1998} determined by eye. \textsuperscript{a} MESSENGER mean speed taken from ENLIL simulation (Sect. 5.2 in Paper I).
}
\end{minipage}
\end{table*}

The locations of MESSENGER, STEREO-A, and STEREO-B in HGI coordinates at the time of the IP shock arrival at STEREO-A (t\textsubscript{S\_STA}) are summarized in Col. 2--4 of Table \ref{tableShockENLILFRONT} and depicted in Fig. \ref{fig:shock_anal}, viewed from the north heliographic pole. In this figure the Sun (yellow) and the Earth (green) are represented with the coloured circles for reference (not to scale). We note that the longitudinal separation between STEREO-A and MESSENGER is $\sim$23$^{\circ}$ at t\textsubscript{S\_STA}, and that MESSENGER is located very close  to the CME apex longitude (orange line in Fig. \ref{fig:shock_anal}).

Assuming radial propagation of the shock, to estimate the global shock surface from in-situ observations based on \cite{Moestl2012}, we plotted in Fig. \ref{fig:shock_anal} the estimated heliocentric distance of the shock (e.g., r\textsubscript{S\_STB} for the shock at STEREO-B) along radials from the Sun to each spacecraft for the time the shock is observed at STEREO-A (t\textsubscript{S\_STA}). For this purpose, we used the relation:

\begin{equation}\label{eq:Shock_distance}
r\textsubscript{S\_STB}=r\textsubscript{STB}+V\textsubscript{S\_STB}(t\textsubscript{S\_STA}-t\textsubscript{S\_STB}), 
\end{equation}

\noindent with V\textsubscript{S\_STB} and t\textsubscript{S\_STB} being respectively the shock speed and its arrival time at STEREO-B, located at a radial distance of r\textsubscript{STB} (as listed in Table \ref{tableShockENLILFRONT}). The result is r\textsubscript{S\_STB}=1.09 au (radial distance indicated with the blue dashed circle in Fig. \ref{fig:shock_anal}), which is larger than r\textsubscript{STA}=0.97 au (red dashed circle). Both distances would be similar if the shock had the shape of a sphere centered on the Sun. 

Similarly, we predicted the distance for MESSENGER. We estimated that the shock speed at MESSENGER ranged from the shock simulated speed value given by ENLIL of V\textsubscript{S\_MESS} = 665 km s$^{\ -1}$ (Sect. 5.2 in Paper I) to the shock transit speed of V\textsubscript{S\_MESS} = 975 km s$^{\ -1}$ (Col. 6 in Table \ref{tableShockENLILFRONT}). Then, the shock distance for MESSENGER (r\textsubscript{S\_MESS}) ranges from 1.00 au to 1.32 au (inner and outer brown dashed circles in Fig. \ref{fig:shock_anal}). These two limit shock distances at MESSENGER point to two different possible shock surfaces, represented by the black bold straight and dashed lines. Between STEREO-B and MESSENGER, and also between STEREO-A and MESSENGER for the outer shock estimated shape, we plotted another point, predicted by eye, to draw a smoother shape of the shock surface.

The normal vector of the shock from the Helsinki IP shock catalogue for STEREO-A (0.88, -0.48, 0.06) and STEREO-B (0.78, -0.56, 0.29), given in RTN coordinates, were translated into Fig. \ref{fig:shock_anal}, as indicated with the arrows next to the spacecraft locations. Magnetic coplanarity calculation at MESSENGER gave a shock normal of (0.30, 0.95, 0.09) in RTN coordinates, indicated with an arrow in Fig. \ref{fig:shock_anal} near the spacecraft. Given the result configuration, near STEREO-B location the shock shape would be closer to the inner r\textsubscript{S\_MESS} (bold font line). Based on the results above, we conclude that the west flank of the IP shock, represented by STEREO-B location, is propagating faster than the other flank (to the east of STEREO-A location). 

We also may assume that the expected $\theta$\textsubscript{Bn} value at 1 au for a radially propagating spherical shock in a nominal magnetic field line configuration is close to 45$^{\circ}$. This is the case for STEREO-B, that observes an oblique IP shock ($\theta$\textsubscript{Bn} = 41$^{\circ}$). However, the IP shock at STEREO-A is quasi-perpendicular ($\theta$\textsubscript{Bn} = 84$^{\circ}$). Thus, STEREO-A is not fulfilling what expected for a spherical shock propagating through a nominal Parker spiral field (Fig. \ref{fig:shock_anal}). This would mean that, in addition to a deviation of the shock normal from the radial direction, the magnetic field lines might be strongly departed from the ideal Parker spiral configuration at STEREO-A location. This might be related with the highly distorted MFR observed at STEREO-A location discussed later. 

In summary, based on our analysis, the IP shock is non-spherical but convex. The west flank, represented by the STEREO-B location, is propagating faster than the east flank, located to the east of STEREO-A (extended dashed black lines in Fig. \ref{fig:shock_anal}). This behaviour is not typical for ICMEs near solar maximum. \cite{Liu2006} found that they were very flattened and convex outward with radii of curvature proportional to their heliospheric distances. However, in this case, the western side of the source region abuts a coronal hole with significant longitudinal extent (Fig. 7 in Paper I) and the western flank of the CME is propagating alongside the high speed stream emanating from that coronal hole. This configuration may explain both the higher shock speed but lower strength and the near oblique shock orientation at STEREO-B. The CME itself is propagating faster along the general direction of MESSENGER and STEREO-A, within the coronagraph field of view. The speed asymmetry is consistent with the expected shock shape represented by dashed lines in Fig. \ref{fig:shock_anal}. 

\subsection{ICME geometry}
\label{sec:ICME_geometry}
 We estimated the geometry of the ICME, such as the longitudinal extent of the ICME and its radial size at the locations of MESSENGER, STEREO-A and STEREO-B. We used the same conventions for the coordinate system and time reference as for the IP shock in the previous section. 

Assuming that both STEREO observed the same ICME and considering it is low tilted with respect to the equatorial plane (|$\tau$| $\leq$ 45{$^{\circ}$}) based on remote-sensing observations (Sect. \ref{sec:Remote-sensing analysis}), the separation between MESSENGER and STEREO-B spacecraft gives an estimate for the half longitudinal extent of the ICME. Such assumption is possible since observations suggest that STEREO-B traversed the ICME close to its western flank and MESSENGER close to its apex as discussed in Sect. \ref{sec:Analysis_IP_shock}. Based on the configuration shown in Fig. \ref{fig:shock_anal}, the longitudinal separation between STEREO-B and MESSENGER is estimated to be 55$^{\circ}$. Thus, the longitudinal extent of the ICME is at least 110$^{\circ}$ ($\sim$1.85 au near 1 au, or $\sim$0.61 au near 0.3 au), which is much higher than the average width at solar maximum of 61$^{\circ}$ \citep{Yashiro2004}. 

The radial diameter of the ICME (D\textsubscript{ICME}) can be estimated by multiplying the ICME duration by the mean speed within the ICME (Col. 10 in Table \ref{table:ICME signtures}) as in \cite{Kilpua2011} or \cite{Jian2006}. This simple analysis does not consider the impact parameter, namely different spacecraft trajectories through the ICME or the closest
approach from the MFR axis. Thus, the obtained radial diameters and aspect ratios (below) should be considered only as rough estimations because several configurations would significantly complicate a simple analysis of the cross-sectional shape. This rough approach gives the following radial diameters: D\textsubscript{ICME\_MESS}=0.36 au, D\textsubscript{ICME\_STB}=0.41 au and D\textsubscript{ICME\_STA}=0.71 au for MESSENGER, STEREO-B and STEREO-A, respectively. 

Thus, the radial extent of the ICME at the location of STEREO-B (D\textsubscript{ICME\_STB}=0.41 au) is within the average ICME size observed by the study carried out by \cite{Jian2006}, who studied 230 ICMEs near 1 au, with a resulting average size of 0.41 ± 0.01 au. However, the radial diameter at the location of STEREO-A (D\textsubscript{ICME\_STA}=0.71 au) is much higher than the average size. This is the same case for MESSENGER, which radial extent (D\textsubscript{ICME\_MESS}=0.36 au) is above the mean size of 0.28 ± 0.01 au analyzed by \cite{2015Winslow}.  

To derive the radial diameter for the MO (D\textsubscript{MO}), as estimated above for the ICME, we used the duration and mean speed within the MO (Col. 13 in Table \ref{table:ICME signtures}). This process resulted in D\textsubscript{MO\_MESS}=0.18 au,
D\textsubscript{MO\_STA}=0.46 au, and D\textsubscript{MO\_STB}=0.11 au for MESSENGER, STEREO-A and STEREO-B, respectively. The difference between radial diameters at the locations of MESSENGER and STEREO-A is mainly related to the expected expansion of the MO between $\sim$0.3 au and $\sim$1 au. As a result of the expansion, the MO size increases with heliocentric distance \citep{Gosling1990}, so that the MO duration (Cols. 4 and 5) at STEREO-A is $\sim$2 days, while it is $\sim$0.5 day at MESSENGER location.

The average speed of expansion (V\textsubscript{exp}) of a MO is defined as a half of the difference between the speed at the LE and trailing-edge (TE) of the MO. For STEREO-A (STEREO-B), the measured speeds at the LE and TE are 457 (601) km s\textsuperscript{-1} and 390 (531) km s\textsuperscript{-1}, respectively, yielding an expansion of V\textsubscript{exp} =33.5 (35) km s\textsuperscript{-1}. These values are slightly above the mean value observed within the MO using 20 years of Wind ICME observations \citep{Nieves-Chinchilla2018DiP}.

We also estimated the aspect ratio of the MO, namely, the ratio of the radial width to the longitudinal extent. The aspect ratio of the MO is $\sim$1:4 at the locations of MESSENGER and STEREO-A, and $\sim$1:16 at STEREO-B. The similar aspect ratios at MESSENGER and STEREO-A might be related with the self-similar expansion of the MO between the two locations. We note that the aspect ratio of the MO is very different at the twin STEREO spacecraft. Although the dynamic solar wind may have varying influence, and the MO could be compressed and disturbed differently at different locations, the main reason for the distinct aspect ratios might be related to different impact parameters at the locations of STEREO-A and STEREO-B, as it will be discussed below.  

The total pressure shown in panel 8 of Fig. \ref{fig:icme_sta_stb} is a parameter that can be used to estimate the closest approach of the spacecraft from the core of the ICME \citep{Russell2005}. At STEREO-B, the total pressure presents a gradual increase followed by a gradual decay, suggesting the ICME is crossed far from the center \citep[Group 3 based on][]{Jian2006}. In contrast, the total pressure at STEREO-A presents a prompt and large increase followed by a half-day plateau (on DOY 235) and then a gradual decay, similar to Group 2 ICMEs in \cite{Jian2006}, suggesting a closer-to-core encounter at STEREO-A. In addition, the sheath region after the shock at STEREO-A is proportionally shorter than at STEREO-B, the total pressure and the magnetic field strength of the MO are stronger at STEREO-A ($\sim$17 nT, Col. 11 in Table \ref{table:ICME signtures}) than at STEREO-B ($\sim$9 nT). Thus, observations suggest that STEREO-A traversed the ICME closer to the core of the ICME than STEREO-B \citep{Kilpua2011}.

As the MO at the locations of MESSENGER and STEREO-A are compounded of two different structures, MC and MC-like, we performed a similar estimation of the radial extent as presented above but only for the MC, with the following results: D\textsubscript{MC\_MESS}=0.07 au and D\textsubscript{MC\_STA}=0.24 au. We compared these values with the average expansion of MCs in the inner heliosphere at any heliocentric distance based on \cite{Leitner2007}. They studied 130 events observed by the Helios 1 and Helios 2 probes between 0.3 and 1 au in the ascending and maximum phases of solar cycle 21 and by Wind \citep{Szabo2015}.  Then, we used:

\begin{equation}\label{eq:ICME_expansion}
D\textsuperscript{'}=D\textsubscript{h}r\textsuperscript{'}/r\textsubscript{h}\textsuperscript{1.14}, 
\end{equation}

\noindent as a rough proxy, where the diameter D\textsubscript{h} of the MC is observed at a heliocentric distance r\textsubscript{h}, with D\textsubscript{h} and r\textsubscript{h} in au. Then, the estimated expansion of the MC diameter measured at MESSENGER when it arrives at the location of STEREO-A (r\textsubscript{MESS}=0.34 au and D\textsubscript{MC\_MESS}=0.07 au), using r\textsuperscript{'}=0.97 au, would be D\textsuperscript{'}=0.23 au, similar to the estimated value of 0.24 au. This suggests that MESSENGER and STEREO-A observe a similar MC structure, likely corresponding to the F2 part of the CME in Fig.~\ref{fig:GCS_recon}.

Finally, we used another proxy based on \cite{Leitner2007}, who found that the maximum magnetic field strength inside MCs in the inner heliosphere (<1 au) depends on radial distance. Extrapolating this relation to the ICME of this study, we scaled the maximum magnetic field strengths (Col. 11 in Table \ref{table:ICME signtures}), observed at the MO for the three locations, to the spacecraft taken as reference (STEREO-A). Then, to compensate for the heliocentric distances r of the spacecraft, we used the relation: 
\begin{equation}\label{eq:Magnetic_strength_distance}
B\textsuperscript{'}\textsubscript{max\_s/c}=B\textsubscript{max\_s/c}(r\textsubscript{STA}/r\textsubscript{s/c})\textsuperscript{-1.64}, 
\end{equation} 

\noindent with B in (nT) and r in (au). The maximum field strengths extrapolated at r\textsubscript{STA} (Col. 12 in Table \ref{table:ICME signtures}) compare as 28.8 nT for MESSENGER, 10.0 nT for STEREO-B, and 16.6 nT for STEREO-A (not scaled). Thus, the high peak field at MESSENGER at the front of the MO (168.9 nT) is scaled down to normal solar wind levels at 1 au. It also means that the ICME has a component toward MESSENGER, that is located close to the nose of the ICME, and STEREO-B might be closer to the flank of the ICME than STEREO-A.

\subsection{ICME magnetic field configuration}
\label{sec:MC magnetic field configuration}

The visual analysis of the magnetic field strength, as observed in the first panels of Fig. \ref{fig:messenger} and Fig. \ref{fig:icme_sta_stb} (left), suggests that the expansion of the ICME between MESSENGER and STEREO-A shifts the magnetic field maximum even further towards the LE, and results in declining speed profiles (left panel 5 of Fig. \ref{fig:icme_sta_stb}). To quantify the asymmetry of the magnetic field strength profile, \cite{Nieves-Chinchilla2018DiP} introduced the distortion parameter (DiP). The declining magnetic field magnitude with magnetic field compression at its front, observed within the MC at the locations of MESSENGER and STEREO-A, respectively yields to DiP\textsubscript{MESS}=0.44 and DiP\textsubscript{STA}=0.42. The lack of plasma observations at MESSENGER, crossing the ICME close to the apex, prevents us from discerning if this is a signature of distortion or expansion. 

The asymmetry in the magnetic field profile observed at STEREO-A (DiP\textsubscript{STA}=0.42) might be interpreted as expansion ($\sim$V\textsubscript{exp}=33.5 km s\textsuperscript{-1}), but could be also explained as evidence of a distorted flux rope, as it is suggested by the pileup feature C in Fig.~\ref{fig:GCS_recon}. There might be also some erosion, observed as magnetic field fluctuations at the front of the MC, as discussed in section \ref{sec:OBSER_in_situ} for MESSENGER. Thus, the presence of distortion in the MC would justify the use of an elliptical cylindrical fitting (EC model) of the flux rope presented in the following section. The DiP parameter calculated within the MC at STEREO-B is DiP\textsubscript{STB}=0.48, meaning a symmetric profile, but with an expansion present (V\textsubscript{exp}=35 km s\textsuperscript{-1}). We note the distinct DiP parameters between STEREO-B and STEREO-A locations.

\begin{figure}[htbp] 
   \resizebox{\hsize}{!}{\includegraphics{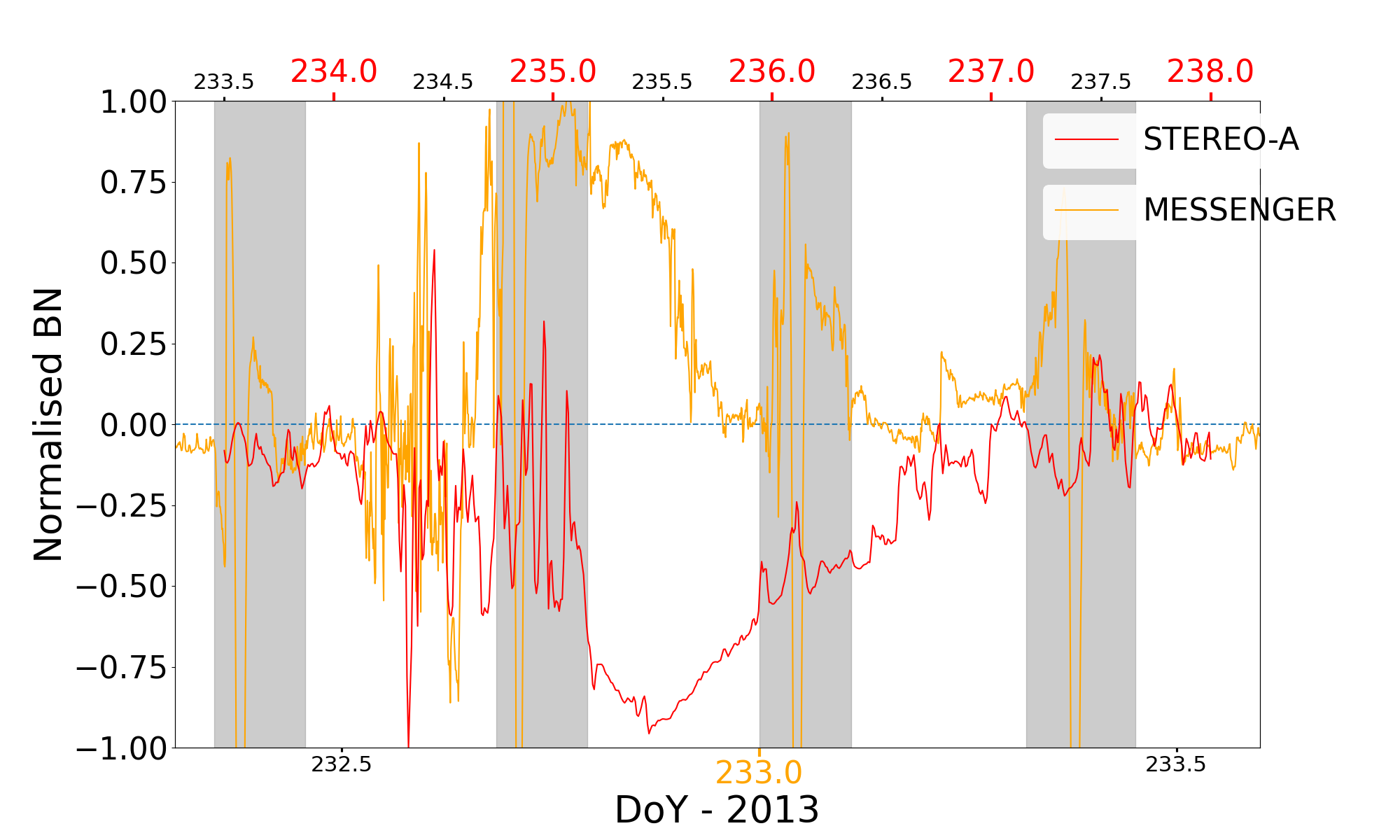}}
     \caption{Comparison of the in-situ magnetic field  B\textsubscript{N} component observed by MESSENGER (orange) and STEREO-A (red). The magnetic field is normalised to their respective absolute maximum value, represented in the same RTN coordinates, with a time shift and data expansion, as the MFR arrives to the spacecraft at different times and has different duration. The upper and lower horizontal axes respectively represent the STEREO-A and MESSENGER observing time. The gray shaded areas indicate MESSENGER crossings through Mercury's magnetosphere.}
     \label{fig:BN}
\end{figure}

 \begin{table*}[htbp]
\caption{EC model fit parameters in RTN coordinates.}
\label{table:ECmodel}
\begin{tabularx}{1\textwidth}{ccccccccc} 
\hline
\hline
s/c\ & Longitude  & Tilt & Rotation &Ellipse ratio  &Cross-section Radius &Distance & $\chi\textsuperscript{2}$& Chirality\\
&$\phi$ (deg)&$\theta$ (deg)&$\xi$ (deg)&$\delta$ (-)&R (au)&$Y_0$ (au)\\
\hline
(1)&(2)&(3)&(4)&(5)&(6)&(7)&(8)&(9)\\

\hline
MESSENGER MC&  295 &  25 &  73 & 0.44 & 0.052 &  -0.019& 0.15&Positive\\
STEREO-B MC&58& -13& 134& 0.68& 0.137&0.109&0.29&Negative\\
STEREO-A MC& 325 & -46 &  79 & 0.67 & 0.160 & -0.086&0.22&Positive \\
 \hline
\\
\end{tabularx}
\footnotesize{\textbf{Notes.} Column 1: Spacecraft. Column 2: MFR axis longitude ($\phi$=[0...360]$^{\circ}$). Column 3: inclination of the flux rope with respect to the equatorial plane ($\theta$=[-90...+90]$^{\circ}$). Column 4: MFR rotation about its central axis  ($\xi$=[{0...180}]$^{\circ}$). Column 5: MFR distortion (ratio between major and minor ellipse axis, $\delta$=[0...1]). Column 6: MFR size. Column 7: distance from the spacecraft trajectory to the MFR axis (negative value means that the spacecraft is crossing the upper part of the structure). Column 8: goodness of the fitting ($\chi\textsuperscript{2}$=[0...1]). Column 9: MFR handedness.
 }
\end{table*}

\begin{figure*}
\centering
   \resizebox{\hsize}{!}{\includegraphics{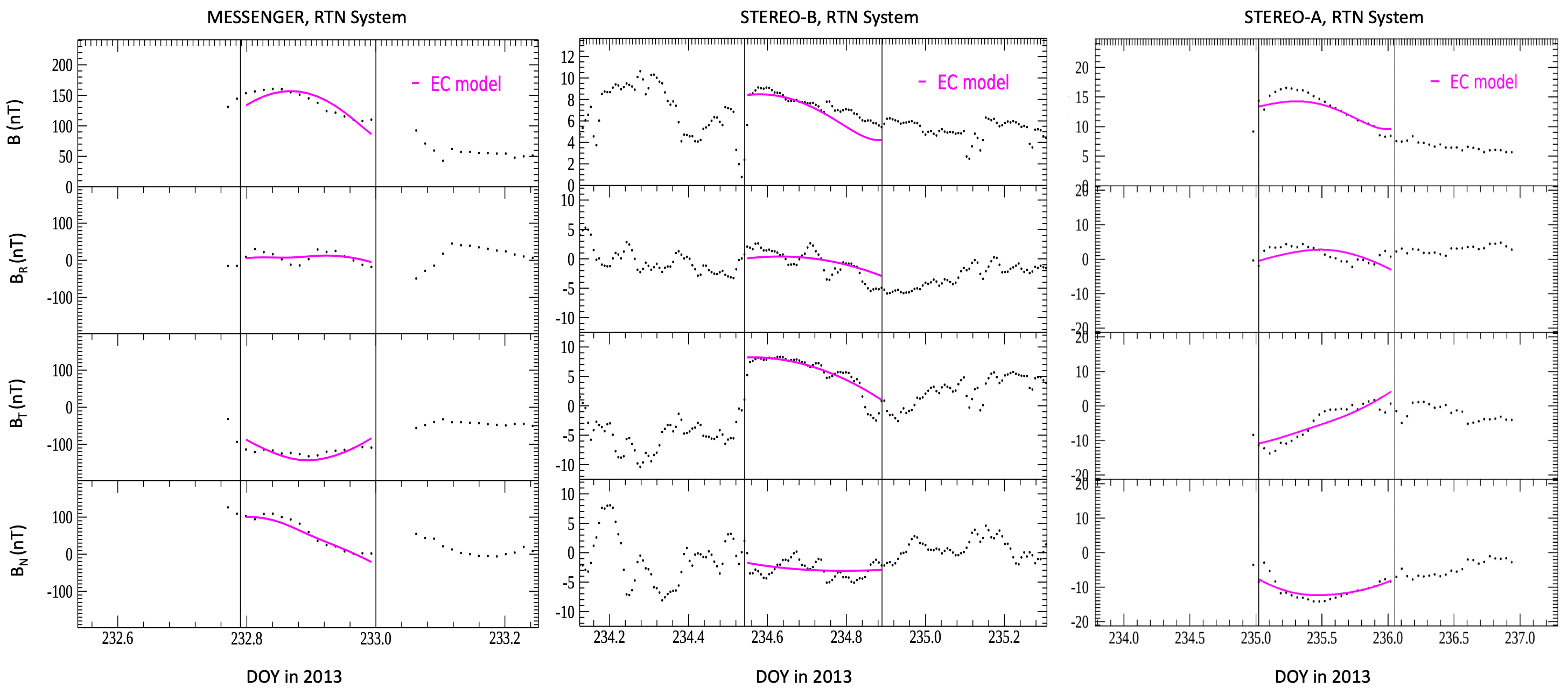}}
     \caption{Comparison of the EC model  fitting results (pink) with MESSENGER (\textit{left}),  STEREO-B (\textit{center}), and STEREO-A (\textit{right})  magnetic field observations spanning the MC. \textit{From the top}, the panels display the magnetic field strength and the three magnetic field B\textsubscript{RTN} components, respectively.}
     \label{fig:EC_model_mag_comp}
\end{figure*}
To classify the MC based on the different magnetic configurations \citep{BothmerSchwenn1998} at the different spacecraft (Col. 14 in Table \ref{table:ICME signtures}), we visually analysed the magnetic field components at each location. MESSENGER (Fig. \ref{fig:messenger}) encountered a MC with a similar configuration as a bipolar one, that is, being low tilted with respect to the ecliptic plane (|$\tau$| $\leq$ 45{$^{\circ}$}). The magnetic field B\textsubscript{N} component (second panel) goes from highly positive to zero values within the MC (blue shading). The magnetic field azimuthal angle, $\theta$\textsubscript{B-RTN} (fourth), is always positive, but showing a steep decrease, and the longitudinal angle, $\phi$\textsubscript{B-RTN}, (3) is pointing to the east. This means that the magnetic field rotates slightly from north to south and at the cloud center the field points eastwards. This configuration corresponds to a NES (North-East-South) configuration, that represents a positive or RH chirality.

The visual analysis of the magnetic field observed by STEREO-A (left panels of Fig. \ref{fig:icme_sta_stb}) derives that the spacecraft encountered an unipolar MC (highly tilted, i.e., |$\tau$| $\geq$ 45{$^{\circ}$}). The magnetic field B\textsubscript{N} component (2) is always negative within the MC (blue shading). The magnetic field azimuthal angle, $\theta$\textsubscript{B-RTN} (4), is always negative, and the longitudinal angle, $\phi$\textsubscript{B-RTN} (3), rotates from positive to negative polarity. This means that the magnetic field rotates smoothly from east to west and at the cloud center the field points southwards. This configuration corresponds to a ESW (East-South-West) cloud type of  positive (RH) chilarity. Thus, having the same helicity sign, the MC orientations observed at MESSENGER and STEREO-A locations are different. Figure \ref{fig:BN} presents the magnetic field B\textsubscript{N} component as observed by MESSENGER and STEREO-A. They are normalized to their respective absolute maximum value (avoiding the magnetospheric crossings) and shifted in time and expanded for an easier comparison. The B\textsubscript{N} component for both spacecraft are similar but with opposite signs.

Finally, the visual analysis of the magnetic field observed by STEREO-B (right panels of Fig. \ref{fig:icme_sta_stb}) derived that the spacecraft encountered a unipolar MC (high tilted, i.e.,  |$\tau$| $\geq$ 45{$^{\circ}$}). The magnetic field B\textsubscript{N} component (2) is always negative within the MC (blue shading). The magnetic field azimuthal angle, $\theta$\textsubscript{B-RTN} (4), keeps pointing south, and the longitudinal angle, $\phi$\textsubscript{B-RTN} (3), is rotating from west to east. This means that the magnetic field rotates smoothly from west to east and at the cloud center the field points southwards. This configuration corresponds to a WSE (West-South-East) cloud type of negative or LH chilarity. We note that the negative helicity observed at STEREO-B is distinct from the positive helicity observed at the locations of MESSENGER and STEREO-A.










\subsection{MC reconstruction}
\label{sec:Modelling_ICME}

The analytical MFR model or EC model, developed by \cite{2018aNievesChinchilla}, was applied to reconstruct the MC present within the ICME at the locations of MESSENGER, STEREO-A, and STEREO-B. The MC reconstructions are local, based on the magnetic field measured in situ at each location. The EC model assumes a MFR magnetic topology, that is, an axially symmetric magnetic field cylinder with twisted magnetic field lines of elliptical cross section. Therefore, the EC model allows us to consider cross-section distortion as a consequence of the interaction of the flux rope with the solar wind, as discussed in Sect. \ref{sec:MC magnetic field configuration}. In Cols. 6 and 7 of Table \ref{table:ICME signtures} we list the MC time intervals chosen for the EC model analysis, corresponding to the blue shadings in Fig. \ref{fig:messenger} (MESSENGER) and Fig. \ref{fig:icme_sta_stb} (STEREO). Column 13 in Table \ref{table:ICME signtures} shows the average solar wind speed used for the fitting. The trajectory of the spacecraft through the MC is inferred by using the minimization of the $\chi\textsuperscript{2}$ function to obtain a set of parameters that best fit the measured data \citep{2018bNieves-Chinchilla}. Table \ref{table:ECmodel} lists the obtained $\chi\textsuperscript{2}$ function and the EC model fit parameters in RTN coordinates. The MFR orientation in space is given by three angles: the central magnetic field longitude, $\phi$ (equal to 0$^{\circ}$ in the spacecraft-Sun direction), the tilt angle, $\theta$ (where positive values represent north of the equatorial plane), and the MFR rotation about its central axis, $\xi$. The geometry of the flux rope is given by the ratio between major and minor ellipse axis, $\delta$, and the size by the cross-section major radius, R. $Y_0$ is the impact parameter, that represents the closest approach to the MFR axis, where a positive value means that the spacecraft is crossing the lower part of the structure. Finally, the chirality or handedness of the flux rope is shown in Col. 9. 

In Fig. \ref{fig:EC_model_mag_comp}, the magnetic field data from MESSENGER (left), STEREO-B (middle) and STEREO-A (right) are shown, along with the EC model fitting (smooth pink lines). Despite the model following the trend of the magnetic field components, the changes in B\textsubscript{R} component are not well captured, especially at the locations of MESSENGER and STEREO-B. The opposite behaviour is found in the reproduction of the B\textsubscript{T} component, where the changes in STEREO-A are not well followed. The B\textsubscript{N} fitting component at STEREO-B does not reproduce the fluctuations of the magnetic field. 

Then, according to Table \ref{table:ECmodel}, MESSENGER observes the MFR axis approximately close to perpendicular to the radial direction, based on the magnetic
field longitude value ($\phi$=295$^{\circ}$), close to 270$^{\circ}$, while STEREO-A, with a longitude angle closer to 360$^{\circ}$ ($\phi$=325$^{\circ}$) might observe the flux rope closer to a flank. In case of STEREO-B, the MFR axis is also observed closer to a flank ($\phi$=58$^{\circ}$). 

The tilt angle ($\theta$) shows a remarkable difference between the observatories, with a northwards tilt in MESSENGER ($\theta$=23$^{\circ}$) and a southwards tilt in STEREO-A ($\theta$=-46$^{\circ}$). STEREO-B presents a smaller southwards tilt ($\theta$=-13$^{\circ}$). The agreement in the MFR rotation about its central axis, $\xi$, for MESSENGER and STEREO-A means that the orientation of the ellipse’s major axis is similar in space, but it is not the case for STEREO-B. In the context of the other two angles, the respective $\xi$ value of 73$^{\circ}$ and 79$^{\circ}$ for MESSENGER and STEREO-A, means that the distorted structure is perpendicular to the spacecraft trajectory, in general agreement with the distortions seen in the coronagraphs images. 

An increase in the radius of the MFR cross-section, R, takes place from MESSENGER to both STEREO, what would be expected due to the radial expansion of the MC. The closest distance to the MFR axis, Y\textsubscript{0}, is negative for both MESSENGER and STEREO-A, so that both spacecraft would be crossing the upper part of the structure. MESSENGER is crossing closer to the MFR axis than STEREO-A. In case of STEREO-B, the positive distance means that the spacecraft might be crossing the lower part of the structure, further away from the MFR axis in comparison with STEREO-A. The chirality for both MESSENGER and STEREO-A cylinders is positive, corresponding to RH flux ropes, and LH (negative) for STEREO-B. The fitting results based on $\chi\textsuperscript{2}$ (Col. 8 in Table \ref{table:ECmodel}) and visual inspection (Fig. \ref{fig:EC_model_mag_comp}) give satisfactory results. 


\begin{figure*}
\centering
    \includegraphics[width=12cm]{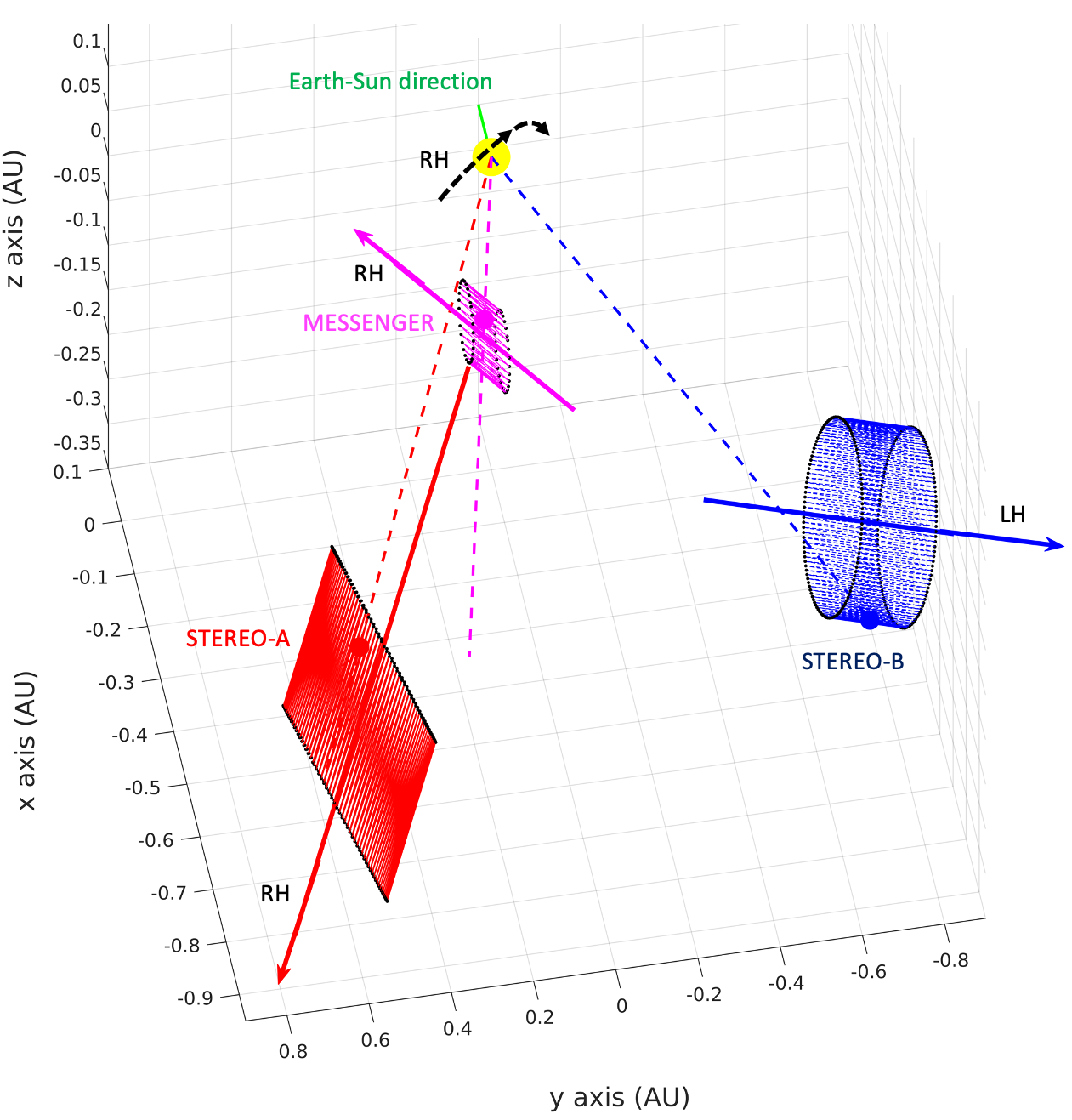}
     \caption{ Global configuration of the magnetic flux ropes. The elliptical cylinders (EC model) representing the MFRs are respectively shown in magenta, red and blue for MESSENGER, STEREO-A and STEREO-B. The respective colored arrows correspond to the cylinders axes final orientation in space. The positions of the spacecraft and the lines connecting them to the Sun are marked with the respective coloured dots (not to scale) and dashed lines. The Sun is indicated with the yellow dot (not to scale) and the green line represents the Earth-Sun direction. The black dashed curve represent the MFR axis orientation as derived from solar disk observations. The helicities as remotely observed and measured in situ are indicated with RH or LH. }
     \label{fig:ICME_reconstruction}
\end{figure*}
\section{Remote/in-situ reconciliation}
\label{sec:remote in situ reconciliation}



The GCS analysis, presented in Sect. \ref{sec:GCS_analysis}, indicates that the western part of the MFR was directed towards W170N13, with a low tilt of $\sim$22$^{\circ}$. Given the estimated half-width of 70$^{\circ}$, the ICME western flank is expected to intercept STEREO-B, located at W222S07. The eastern part of the MFR was directed towards W160S10, with a higher tilt of $\sim$36$^{\circ}$. Based on the half-width of 39$^{\circ}$ of this eastern part, STEREO-A, located at W144S05, is expected to intercept the upper part of the MFR, between the apex and the east flank side. MESSENGER, located at W159N01, is also expected to intercept the upper part of the MFR but closer to the front area and closer to the MFR axis than STEREO-A. Fig. \ref{fig:ICME_reconstruction} shows the configuration of the MFRs, as estimated by solar disk observations and by the geometrical interpretation of the EC model fit parameters presented in Table \ref{table:ECmodel} for each in-situ location.

As derived from solar disk observations (Sect. \ref{sec:CME_flux_rope_type}), the MFR has an undulating axis with a tilt varying between $\sim$36$^{\circ}$ in the east to $\sim$22$^{\circ}$ in the west, and oriented to the west, corresponding to a positive chirality (RH). This imaging-based MFR axis orientation is indicated with the black dashed curved arrow in Fig. \ref{fig:ICME_reconstruction}. The MFR axis orientation from the MESSENGER in-situ reconstruction  (magenta)  is oriented to the northeast (magenta arrow) and hence it is inconsistent to the imaging-based axis. Similarly for STEREO-A, the in-situ reconstructed MFR (red) is oriented to the southeast (red arrow) while the MFR axis as derived from solar disk observations is oriented to the northwest. Possible explanations for these discrepancies are discussed in the following section. As predicted by the GCS reconstruction, MESSENGER and STEREO-A are observing the upper part of the structure, with MESSENGER (magenta point) crossing closer to the MFR axis than STEREO-A (red point). STEREO-B exhibits a different behaviour than MESSENGER and STEREO-A, as both the tilt and orientation of the MFR axis (blue cylinder) are similar to the image-based MFR values. STEREO-B (blue point) is likely intercepting  the lower part of the structure far from the core. 

Regarding the helicities, the intrinsic MFR type analysis using solar disk observations derived a positive helicity (RH). The visual analysis of in-situ data and EC model also indicated a positive helicity for MESSENGER and STEREO-A, but a negative helicity (LH) for STEREO-B.
As discussed in the following section, this could be related to MESSENGER and STEREO-A observing the core of the MFR and STEREO-B observing one leg away from the MFR core.  

The estimated width of the CME based on GCS fitting was 119$^{\circ}$ at 13 R\textsubscript{$\odot$}, very similar to the longitudinal extension of 110$^{\circ}$ of the ICME at 1 au estimated from in-situ data. From the in-situ analysis, the ICME arrived near frontally to MESSENGER, close to the west flank at STEREO-B, and closer to the east flank than to the nose at STEREO-A. The ICME western flank (STEREO-B) arrived earlier than the east flank at locations near 1 au. As discussed above, this could be due to the proximity of a coronal hole (and the resulting fast wind) along the direction of the CME propagation towards STEREO-B. 

\section{Discussion}
\label{sec:summary and discussion}


On 2013 August 19, a CME was ejected from an AR located near the far-side central meridian from Earth's perspective. The CME-driven coronal shock was associated with an unusual large widespread SEP event discussed in detail in \cite{2021Rodriguez-Garcia}. The CME was wide, with an equatorial extent of $\sim$119$^{\circ}$, and the MFR axis presented a curved shape, with two different tilt angles in the west and east sides, showing a positive (RH) helicity. The corresponding ICME was fast enough to drive an IP shock, and the ICME, including the IP shock, was observed in situ by MESSENGER, near 0.3 au, and by STEREO-A and STEREO-B, near 1 au, spanning a longitude of 78$^{\circ}$ in the heliosphere. 


The first spacecraft observing the ICME was MESSENGER, where the IP shock and MO respectively arrived at DOY 232.53 and 232.79 in 2013. A first MC was observed within the MO, followed by a MC-like structure. Thus, the magnetic structure as observed by MESSENGER is classified as complex. The ICME was observed near frontally and very close to the core of the structure, in agreement with the relative position between the CME apex (W165) and MESSENGER location (W159) and with the different hints observed in situ. For example, the relative short duration of the sheath region in comparison to the MO duration ($\sim$5 hours versus $\sim$12 hours), the effect of the ICME arrival in the SEP profile (sudden decrease of near-relativistic electrons), or the magnetic field strength value in the MO of $\sim$170 nT. This increased magnetic field value measured at MESSENGER might be related with the compression area within the CME part that was oriented towards MESSENGER, as observed by the coronagraph images (pileup feature C in Fig. \ref{fig:GCS_recon}). The EC model also estimated a near frontal arrival, with the spacecraft passing through the upper part of the MFR, very near to the central axis. This is in agreement with the orientation of the eastern part (F2) of the MFR erupting from the Sun and the longitude and latitude coordinates of MESSENGER. The MFR type based on in-situ data visual analysis and on the EC model corresponded to a low-tilted cloud with positive helicity, the same observed at the Sun. However, the orientation of the in-situ cylinder (north east) is inconsistent with the MFR axis derived from the solar disk observations (north west). This discrepancy could be related to the visual selection of the MC boundaries, since no solar wind plasma was available. However, a somewhat similar discrepancy exists at STEREO-A location, where we are confident about the limits chosen for the MC.   

STEREO-A, located 15$^{\circ}$ east from MESSENGER, observed the IP shock and MO respectively arriving at DOY 234.29 and 234.97 in 2013. STEREO-A also observed two different MFR structures within the MO, a MC and MC-like. Thus, the magnetic structure is also classified as complex at STEREO-A location. Supported by the relative position of STEREO-A and the direction of the east side of the CME eruption, STEREO-A is observing the eastern part (F2) of the ICME, close to the MFR axis, but closer to the flank area than to the CME nose. Several in-situ proxies support this scenario. The transient depression observed in protons up to 30 MeV might be related to STEREO-A measuring close to the core of the closed magnetic field topology of the structure. However, the maximum magnetic field observed at MESSENGER scaled down to 1 au (28.8 nT) is much higher than the one observed at the location of STEREO-A (16.6 nT), suggesting that MESSENGER is much closer to the CME apex than STEREO-A, as expected from Fig. \ref{fig:shock_anal}. The radial extent of the ICME at STEREO-A location (D\textsubscript{ICME\_STA}=0.71 au) is much higher than the average size \citep{Jian2006} also suggesting an oblique arrival. The MFR type based on in-situ data analysis and on the EC model derived a high-tilted ($\sim$46$^{\circ}$) cloud type with positive helicity for STEREO-A. Based on the EC model, the tilt of the MFR is similar (within the uncertainty of $\sim$10$^{\circ}$) to the eastern part of the MFR axis (tilt of $\sim$36$^{\circ}$) based on solar disk observations. However, the orientation of the MFR axis is the opposite, northwest at the Sun and southeast at STEREO-A location.

A possible explanation for the discrepancies in the MFR axis orientation between MESSENGER and STEREO-A, and between both in-situ locations and the orientation estimated at the Sun, is that we are measuring a curved and rather complex MFR CME topology with single trajectories within the ICME. Further to the different heliocentric distance, the relative angular separation between MESSENGER and STEREO-A is also relevant, as the spacecraft measures local plasma properties. Latitudinal variations could also affect the ICME structure. 

However, there is a plausible alternative scenario for these discrepancies, as the distortion and deformation of the MOs associated to the CMEs are not well understood yet. In this study, we have observed a distorted CME front directed to MESSENGER and STEREO-A, based on the coranagraph images (pileup features C and F in Fig. \ref{fig:GCS_recon}); and we have measured in situ a highly asymmetric magnetic field profile in the magnetic strength within a complex structure in both locations, where the estimated DiP parameters for MESSENGER and STEREO-A respectively were 0.44 and 0.42. Thus, the expansion of the ICME between MESSENGER and STEREO-A shifts the magnetic field maximum even further towards the LE. To carry out the study, we have identified, classified and studied the structures according to the current understanding of the flux rope internal structure. However, either the boundaries selection or the model-reconstruction technique used in the analysis could not describe the actual MO. We have separated the MOs in two structures, since there is not a model able to reproduce the compounded structure (MC and MC-like). The second assumption is the selection of the elliptical-cylindrical cross-section model as the proper one to describe these structures. Now, thanks to the multipoint observations, we observe these discrepancies between MESSENGER and STEREO-A measurements, which may be better described with models including more complex distortions.  



Thus, MESSENGER and STEREO-A could be measuring the same part of a curved MFR, with MESSENGER measuring very close to the CME apex, showing a complex structure, and both spacecraft presenting highly distorted fronts that could change the magnetic field components in such manner that might affect the final orientation of the MFR axis derived from them.

STEREO-B presented a different behaviour than MESSENGER and STEREO-A. Although it was located 0.05 au further from the Sun than STEREO-A, STEREO-B observed the IP shock arriving at 234.09, five hours earlier than STEREO-A. This might be due to the western part (F1) of the CME propagating along a coronal hole with significant longitudinal extent that may explain the higher shock speed. In addition, the higher solar wind speed and lower density observed at the location of STEREO-B, explain the higher speed of the ICME shock west flank. This is consistent with the west flank of the ICME  propagating faster than the east flank.

The MO arrived at STEREO-B at DOY 234.54 in 2013, where the sheath region was wide, but there were signatures, in terms of magnetic field rotation, of the MC being observed by STEREO-B during eight hours. This short period of MC observation is related with STEREO-B encountering the ICME by a flank and far away from the MFR axis. This is in agreement with the orientation of the western part of the CME and the longitude and latitude coordinates of STEREO-B. Based on the EC model, STEREO-B cylinder axis is oriented to the west and it is observing the lower part of the structure further away from the MFR axis, consistent with STEREO-B observing the western leg of the MFR. Supporting this scenario, the SEP profile at STEREO-B is not influenced by the ICME arrival, so the spacecraft is crossing far from the closed topology of the structure. Regarding the MC helicity, based on in-situ data visual analysis and EC model fitting, it was negative, opposite to MESSENGER and STEREO-A.

The plausible reasons for this difference in helicity are related with STEREO-B observing one leg of the MFR far from the core, while MESSENGER and STEREO-A are observing closer to center and to the core of the structure. Firstly, as based on \cite{Mulligan1999}, the helicity remains the same for the same MFRs during the evolution in the heliosphere. There are previous studies that support the possibility of magnetic flux added in opposite directions during the first stages of the CME evolution \citep{2013Cho,2017VemareddyDemoulin}, so the center and the legs of the flux rope can have opposite helicities. This would have a destabilizing effect on the ICME. At this respect, there are theoretical analysis on the helical kink instability and the reversed chirality scenario \citep{Florido2020}, that could be part of a future further investigation. In addition, the simulation of the flux emergence to replicate the magnetic field distribution would be also interesting to study the dual chirality at the eruption, but it is out of the scope of this paper.  Secondly, the considerable crossing distance of STEREO-B from the MFR axis, could lead us to an erroneous helicity estimation. The limited observations of the MFR make the methods (both visual inspection and EC model), uncertain. In addition, the magnetic reconnection related to the erosion present in the MFR, probably due to the proximity of the HCS, could also influence the estimated helicity \citep[][and references therein]{2021Pal}. 


Thus, MESSENGER, STEREO-A and STEREO-B are observing the same ICME. This means that the longitudinal extent of the ICME is at least 78$^{\circ}$ near 1 au, with an estimated extent of 110$^{\circ}$, which is much higher than solar maximum average of 61$^{\circ}$ \citep{Yashiro2004}.

\section{Conclusions}
\label{sec:conclusions}

Our main conclusions can be summarized as follows:
\begin{description}
 
 \item[$\bullet$ Source region evolution:] The multi-viewpoint EUV observations revealed a fan-like eruption, spreading from west to east while propagating outwards. Despite the lack of photospheric magnetic field information for this far-side eruption, the morphology and evolution of the coronal loops images by STEREO, provided detailed information for the extent, helicity and ambient conditions of the eruption.
 
 \item[$\bullet$ Coronal evolution:] The multi-viewpoint observations were instrumental in understanding the white light CME. The significant longitudinal expansion, seen in the EUV images, gave rise to asymmetric white light signatures. Without coronagraph observations from  three viewpoints (STEREO-A, STEREO-B, and LASCO), the event could have easily been interpreted as two separate CMEs, resulting in a misguided analysis. As it turns out, we were able to reproduce the complexity and large longitudinal extent of the event via two separate reconstructions of the observed white light signatures. The resulting kinematic and dynamical properties of the CME were consistent with the EUV source region and in-situ signatures.

\item[$\bullet$ ICME radial evolution:] The disagreement in the MFR orientation observed at the solar disk and measured at MESSENGER, located at 0.33 au, and STEREO-A, near 1 au, is due to measuring a curved, highly distorted and rather complex MFR magnetic topology. 

\item[$\bullet$ Evidence of a complex structure:] At the locations of MESSENGER and STEREO-A, the magnetic field structure might be complex, composed of a MC (full rotation) followed by a MC-like (partial rotation).

\item[$\bullet$ ICME longitudinal evolution I:] The MFR observed at the location of STEREO-B, which presented a negative helicity, and the MFR observed at the locations of MESSENGER and STEREO-A, both with positive helicity, belong to the same magnetic structure. The opposite helicity might be related to STEREO-B observing the western leg of the ICME far from the core, while MESSENGER and STEREO-A are observing the core of it, closer to the ICME apex.

 
 \item[$\bullet$ ICME longitudinal evolution II:] The longitudinal extent of the ICME is 110$^{\circ}$ near 1 au. Based on in-situ data, the IP shock west flank (STEREO-B) travels faster and has an earlier arrival than the east flank of the ICME at 1 au. The CME propagating against a coronal hole and different local plasma conditions in which the ICME propagates may have caused distortion in the ICME shape.

\end{description}


This work illustrates how a wide, curved, highly distorted and rather complex CME was showing different  orientations as observed on the solar disk and measured in situ at 0.3 au and near 1 au. It also shows how the ambient conditions, namely the presence of a coronal hole and corresponding high speed stream (along the western edge of the eruption) or propagation along the dense heliospheric plasma sheet (pileup features C and F in Figure~\ref{fig:GCS_recon}), can significantly affect the expansion and propagation of the CME/ICME, introducing additional irregularities to the already asymmetric eruption. New missions like Solar Orbiter \citep{Muller2020,Zouganelis2020}, launched in February 2020, or Parker Solar Probe \citep[][]{Fox2016}, launched in August 2018, are providing observations of unexplored areas where the pristine flux ropes are less affected by various phenomena throughout their evolution in the heliosphere. With the unprecedented high-resolution remote-sensing instruments and sampling more regions in the inner heliosphere, together with spacecraft near 1 au, we could discern if we are observing local or global orientations of the flux ropes and what would be the most probable scenarios for observing complex structures and different orientations in the heliosphere.  



\begin{acknowledgements}
The UAH team acknowledges the financial support by the Spanish Ministerio de Ciencia, Innovación y Universidades FEDER/MCIU/AEI Projects ESP2017-88436-R and PID2019-104863RB-I00/AEI/10.13039/501100011033. LRG is also supported by the European Space Agency, under the ESA/NPI program. LAB acknowledges the support from the NASA program NNH17ZDA001N-LWS (Awards Nr. 80NSSC19K0069 and 80NSSC19K1235). AV was supported by NASA LWS 80NSSC19K0069. MD acknowledges support by the Croatian Science Foundation under the project IP-2020-02-9893 (ICOHOSS). LKJ is supported by NASA’s LWS and HSR programs. FCM acknowledges the financial support by the Spanish MINECO-FPI-2016 with FSE predoctoral grant. LFGS was supported by the NASA Grant No. 80NSSC20K1580. The authors acknowledge the different SOHO, STEREO and MESSENGER instrument teams, and the STEREO science center for providing the data used in this paper. MESSENGER data were downloaded from the Planetary Data System. In addition, this paper uses data from the Heliospheric Shock Database, generated and maintained by the University of Helsinki. Part of the information used in the ICME reconstruction model was presented in 2019, in the Bachelor's Thesis with title "Analysis of Interplanetary Coronal Mass Ejections Observed by STEREO and MESSENGER", of Marina González Álvarez at Universitat Politècnica de Catalunya in collaboration with NASA (\url{https://upcommons.upc.edu/handle/2117/168746}).  

\end{acknowledgements}

\begin{flushleft}

\textbf{ORCID iDs} 
\vspace{2mm}

Laura Rodríguez-García \orcid{https://orcid.org/0000-0003-2361-5510}

Teresa Nieves-Chinchilla
\orcid{https://orcid.org/0000-0003-0565-4890}

Raúl Gómez-Herrero \orcid{https://orcid.org/0000-0002-5705-9236}

Ioannis Zouganelis
\orcid{https://orcid.org/0000-0003-2672-9249}

Angelos Vourlidas
\orcid{https://orcid.org/0000-0002-8164-5948}

Laura Balmaceda \orcid{https://orcid.org/0000-0003-1162-5498}

Mateja Dumbovi\'c
\orcid{https://orcid:0000-0002-8680-8267}

Lan Jian \orcid{https://orcid.org/0000-0002-6849-5527}

Leila Mays \orcid{http://orcid.org/0000-0001-9177-8405}

Fernando Carcaboso Morales
\orcid{https://orcid.org/0000-0003-1758-6194}

Luiz Fernando Guedes dos Santos
\orcid{https://orcid.org/0000-0001-5190-442X}

Javier Rodríguez-Pacheco
\orcid{https://orcid.org/0000-0002-4240-1115}

\end{flushleft}
%
%

\end{document}